\pdfoutput=1

\documentclass[11pt,twoside,a4paper,cmspaper,final,collab]{cms-tdr}
\def\svnVersion{480367P}\def\cmsCernNoTag{CERN-EP-2018-063}\def\cmsCernDate{\today}\def\cmsMessage{Published in the European Physical Journal C as \href{http://dx.doi.org/10.1140/epjc/s10052-018-6332-9}{\doi{10.1140/epjc/s10052-018-6332-9.}}}

\begin{document}\cmsNoteHeader{TOP-17-007}

\hyphenation{had-ron-i-za-tion}
\hyphenation{cal-or-i-me-ter}
\hyphenation{de-vices}
\RCS$HeadURL: svn+ssh://svn.cern.ch/reps/tdr2/papers/TOP-17-007/trunk/TOP-17-007.tex $
\RCS$Id: TOP-17-007.tex 473157 2018-08-28 11:36:41Z stadie $

\newlength\cmsFigWidth
\ifthenelse{\boolean{cms@external}}{\setlength\cmsFigWidth{0.85\columnwidth}}{\setlength\cmsFigWidth{0.4\textwidth}}
\ifthenelse{\boolean{cms@external}}{\providecommand{\cmsLeft}{top\xspace}}{\providecommand{\cmsLeft}{left\xspace}}
\ifthenelse{\boolean{cms@external}}{\providecommand{\cmsRight}{bottom\xspace}}{\providecommand{\cmsRight}{right\xspace}}

\newlength\cmsTabSkip\setlength{\cmsTabSkip}{1ex}

\newcommand{\mtop}{\ensuremath{m_{\PQt}}\xspace}
\newcommand{\qqbarpr}{\ensuremath{\PQq\PAQq^\prime\xspace}}
\newcommand{\figdiff}[3]{
  \begin{figure}[h]
    \includegraphics[width=0.49\textwidth]{diff_#1_MTH_cal.pdf}
    \caption{\diffcaption{#2}{#3} }
    \label{fig:#1}
  \end{figure}
}

\newcommand{\diffcaption}{The filled circles represent the data, and the other symbols are for the simulations. For reasons of clarity, the horizontal bars indicating the bin widths are shown only for the data points and each of the simulations is shown as a single offset point with a vertical error bar representing its statistical uncertainty. The statistical uncertainty of the data is displayed by the inner error bars. For the outer error bars, the systematic uncertainties are added in quadrature.}
\newcommand{\statJSF}{\ensuremath{\,\text{(stat+JSF)}}\xspace}

\cmsNoteHeader{TOP-17-007}

\title{Measurement of the top quark mass with lepton+jets final states using {\Pp\Pp} collisions at $\sqrt{s}=13~\text{TeV}$}

\date{\today}

\abstract{
The mass of the top quark is measured using a sample of \ttbar events collected by the CMS detector using proton-proton collisions at $\sqrt{s}=13$\TeV at the CERN LHC.
Events are selected  with one isolated muon or electron and at least four jets from data corresponding to an integrated luminosity of 35.9\fbinv.
For each event the mass is reconstructed from a kinematic fit of the decay products to a \ttbar hypothesis.
Using the ideogram method, the top quark mass is determined simultaneously with an overall jet energy scale factor (JSF), constrained by the mass of the W boson in \qqbarpr{} decays.
The measurement is calibrated on samples simulated at next-to-leading order matched to a leading-order parton shower.
The top quark mass is found to be $172.25\pm 0.08\statJSF \pm 0.62\syst\GeV$.
The dependence of this result on the kinematic properties of the event is studied and compared to predictions of different models of \ttbar production, and no indications of a bias in the measurements are observed.
}

\hypersetup{
pdfauthor={CMS Collaboration},
pdftitle={Measurement of the top quark mass with lepton+jets final states using pp collisions at sqrt(s)=13 TeV},
pdfsubject={CMS},
pdfkeywords={CMS, physics, top quark, top quark mass}
}
\maketitle

\section{Introduction}
The top quark plays a key role in precision measurements of the standard model (SM) because of its large Yukawa coupling to the Higgs boson.
Top quark loops provide the dominant contribution to radiative corrections to the Higgs boson mass, and accurate measurements of both the top quark mass ($\mtop$) and the Higgs boson mass allow  consistency tests of the SM~\cite{Baak:2014ora}.
In addition,  the  decision whether the SM vacuum is stable or meta-stable needs a precise measurement of $\mtop$ as the Higgs boson quartic coupling at the Planck scale depends heavily on $\mtop$~\cite{Degrassi:2012ry}.

The mass of the top quark has been measured with increasing precision using the invariant mass of different combinations of its decay products~\cite{Olive:2016xmw}.
The measurements by the Tevatron collaborations lead to a combined value of $\mtop=174.30\pm0.65\GeV$~\cite{TevatronElectroweakWorkingGroup:2016lid}, while the ATLAS and CMS Collaborations measured $\mtop=172.84\pm0.70\GeV$~\cite{Aaboud:2016igd} and $\mtop=172.44\pm0.49\GeV$~\cite{Khachatryan:2015hba}, respectively, from the combination of their most precise results.
In parallel, the theoretical interpretation of the measurements and the uncertainties in the measured top quark mass derived from the modeling of the selected variables has significantly improved ~\cite{Hoang:2008xm, Moch:2014tta, Marquard:2015qpa, Beneke:2016cbu, Butenschoen:2016lpz, Hoang:2017btd, Nason:2018qkf}.

Since the publication of the CMS measurements~\cite{Khachatryan:2015hba} for proton-proton (\Pp\Pp) collisions at center-of-mass energies of 7 and 8\TeV (Run~1), new theoretical models have become available and a data set has been collected at $\sqrt{s}=13$\TeV that is larger than the Run~1 data set.
At this higher center-of-mass energy, new data and simulated samples are available for this analysis. The method closely follows the strategy of the most precise CMS Run~1 measurement~\cite{Khachatryan:2015hba}.
While the selected final state, the kinematic reconstruction, and mass extraction technique have not changed, the new simulations describe the data better and allow a more refined estimation of the modeling uncertainties.
In contrast to the Run~1 analysis, the renormalization and factorization scales in the matrix-element (ME) calculation and the scales in the initial- and final-state parton showers (PS) are now varied separately for the evaluation of systematic effects. In addition, we evaluate the impact of different models of color reconnection that were not available for the Run~1 measurements.

The pair-produced top quarks (\ttbar) are assumed to decay weakly into \PW{} bosons and bottom (\cPqb) quarks via $\cPqt \to \cPqb\PW$, with one \PW{} boson decaying into a muon or electron  and its neutrino, and the other into a quark-antiquark (\qqbarpr) pair.
Hence, the minimal final state consists of a muon or electron, at least four jets, and one undetected neutrino. This includes events where a muon or electron from a $\tau$ lepton decay passes the selection criteria.
The analysis employs a kinematic fit of the  decay products to a \ttbar hypothesis and two-dimensional likelihood functions for each event to estimate simultaneously the top quark mass and a scale factor (JSF) to be applied to the momenta of all jets.
The invariant mass of the two jets associated with the  $\PW  \to \qqbarpr$ decay serves as an observable in the likelihood functions to estimate the JSF directly, exploiting the precise knowledge of the \PW{} boson mass from previous measurements~\cite{Olive:2016xmw}.
The analysis is performed on the  data sample collected in 2016 and includes studies of the dependence of the measured mass value on the kinematic properties of the events.

\section{The CMS detector and event reconstruction}
\label{sec:detector}
The central feature of the CMS apparatus is a superconducting solenoid of 6\unit{m} internal diameter, providing a magnetic field of 3.8\unit{T}.
Within the solenoid volume are a silicon pixel and strip tracker, a lead tungstate crystal electromagnetic calorimeter (ECAL), and a brass and scintillator hadron calorimeter (HCAL), each composed of a barrel and two endcap sections.
Forward calorimeters extend the pseudorapidity ($\eta$) coverage provided by the barrel and endcap detectors. Muons are detected in gas-ionization chambers embedded in the steel flux-return yoke outside the solenoid.
A more detailed description of the CMS detector, together with a definition of the coordinate system used and the relevant kinematic variables, can be found in Ref.~\cite{Chatrchyan:2008zzk}.

{\tolerance 600
The particle-flow event algorithm~\cite{Sirunyan:2017ulk} reconstructs and identifies each individual particle with an optimized combination of information from the various elements of the CMS detector. The energy of photons is directly obtained from the ECAL measurement, corrected for zero-suppression effects. The energy of electrons is determined from a combination of the electron momentum at the primary interaction vertex as determined by the tracker, the energy of the corresponding ECAL cluster, and the energy sum of all bremsstrahlung photons spatially compatible with originating from the electron track. The energy of muons is obtained from the curvature of the corresponding track. The energy of charged hadrons is determined from a combination of their momentum measured in the tracker and the matching ECAL and HCAL energy deposits, corrected for zero-suppression effects and for the response function of the calorimeters to hadronic showers. Finally, the energy of neutral hadrons is obtained from the corresponding corrected ECAL and HCAL energy.
\par}

The missing transverse momentum \ptvecmiss is calculated as the negative of the vectorial sum of transverse momenta (\pt) of all particle-flow objects in the event.
Jets are clustered from particle-flow objects using the anti-\kt algorithm with a distance parameter of 0.4~\cite{Cacciari:2005hq,Cacciari:2008gp,Cacciari:2011ma}.
The jet momentum is determined as the vectorial sum of all particle momenta in the jet, and is found from simulation to be within 5 to 10\% of the true momentum over the whole \pt spectrum and detector acceptance.
An offset correction is applied to jet energies to take into account the contribution from additional \Pp\Pp{} interactions within the same or nearby bunch crossings (pileup)~\cite{Cacciari:2007fd}.
All jets are corrected by jet energy corrections (JECs)  based on simulations. Residual JECs which are derived from the energy balance in  $\gamma$/\PZ{} boson + jet, dijet, and multijet events~\cite{Khachatryan:2016kdb} are applied to the jets in data.
The JECs are also propagated to improve the measurement of \ptvecmiss.
The reconstructed vertex with the largest value of summed physics-object $\pt^2$ is taken to be the primary $\Pp\Pp$ interaction vertex. The physics objects chosen are those that have been defined using information from the tracking detector, including jets, \ptvecmiss, and charged leptons.
Additional selection criteria are applied to each event to remove spurious jet-like features originating from isolated noise patterns in certain HCAL regions~\cite{CMS-PAS-JME-16-003}.

\section{Data samples, event generation, and selection}
\label{sec:samples}
The data sample collected with the CMS detector during 2016 at a center-of-mass energy $\sqrt{s} = 13$\TeV has been analyzed.
This corresponds to an integrated luminosity of $35.9 \pm 0.9$\fbinv~\cite{CMS-PAS-LUM-17-001}.
Events are required to pass a single-muon trigger with a minimum threshold on the \pt of an isolated muon of 24\GeV or a single-electron trigger with a \pt threshold for isolated electrons of 32\GeV.

{\tolerance 400
Simulated \ttbar signal events are generated at next-to-leading order (NLO) with \POWHEG v2~\cite{Nason:2004rx,Frixione:2007vw,Alioli:2010xd,Campbell:2014kua} and the \PYTHIA~8.219 PS generator~\cite{Sjostrand:2007gs} using the {CUETP8M2T4} tune~\cite{Skands:2014pea, CMS-PAS-TOP-16-021} for seven different top quark mass values of 166.5, 169.5, 171.5, 172.5, 173.5, 175.5, and  178.5\GeV.
The single top quark background is also simulated using \POWHEG~v2~\cite{Alioli:2009je, Re:2010bp} interfaced with \PYTHIA~8.
The background stemming from single vector boson production is generated at leading order (LO) or NLO with \MGvATNLO v2.2.2~\cite{Alwall:2014hca} matched to the \PYTHIA~8 PS using the \textsc{MLM} prescription~\cite{Alwall:2007fs} for \PW{}+jets and the \textsc{FxFx} prescription~\cite{Frederix:2012ps} for \PZ{}+jets, respectively.
Finally, diboson (\PW\PW, \PW\PZ, and \PZ{}\PZ) and multijet events from quantum chromodynamics (QCD) processes are generated with \PYTHIA~8 for ME generation, PS simulation, and hadronization.
These background samples use the \PYTHIA~8 tune \textsc{CUETP8M1}.
The parton distribution function (PDF) set NNPDF3.0 NLO derived with the strong coupling strength $\alpS=0.118$~\cite{Ball:2014uwa} and its corresponding LO version are used as the default parametrization of the PDFs in all simulations, respectively.
The samples are normalized to the theoretical predictions described in Refs.~\cite{Czakon:2011xx,Li:2012wna,Kant:2014oha,Kidonakis:2012rm,Sjostrand:2007gs}.
All events are further processed by a full simulation of the CMS detector based on \GEANTfour~\cite{Agostinelli:2002hh}.
The simulation includes effects of  pileup with the same multiplicity distribution as in data.
The response and the resolution of simulated jets is corrected to match the data~\cite{Khachatryan:2016kdb}.
\par}

We select events that have exactly one isolated muon with $\pt>26\GeV$ and $\abs{\eta} <2.4$ or exactly one isolated electron with $\pt>34\GeV$ and $\abs{\eta}<2.1$~\cite{Chatrchyan:2012xi, Khachatryan:2015hwa}.
The isolation of a lepton candidate from nearby jet activity is evaluated from the sum of the pileup-corrected \pt of neutral hadrons, charged hadrons, and photon PF candidates within a cone of $\Delta R = \sqrt{\smash[b]{(\Delta\eta)^2+(\Delta\phi)^2}} = 0.4$ for muons and $\Delta R = 0.3$ for electrons.
Here $\Delta\eta$ and $\Delta\phi$ are the differences in the pseudorapidity and azimuthal angles (in radians) between the particles and the lepton candidate.
The sum of the \pt of the particles is required to be less than 15\% of the muon \pt and 10\% of the electron \pt, respectively.

\begin{figure*}[!tb]
\includegraphics[width=0.49\textwidth]{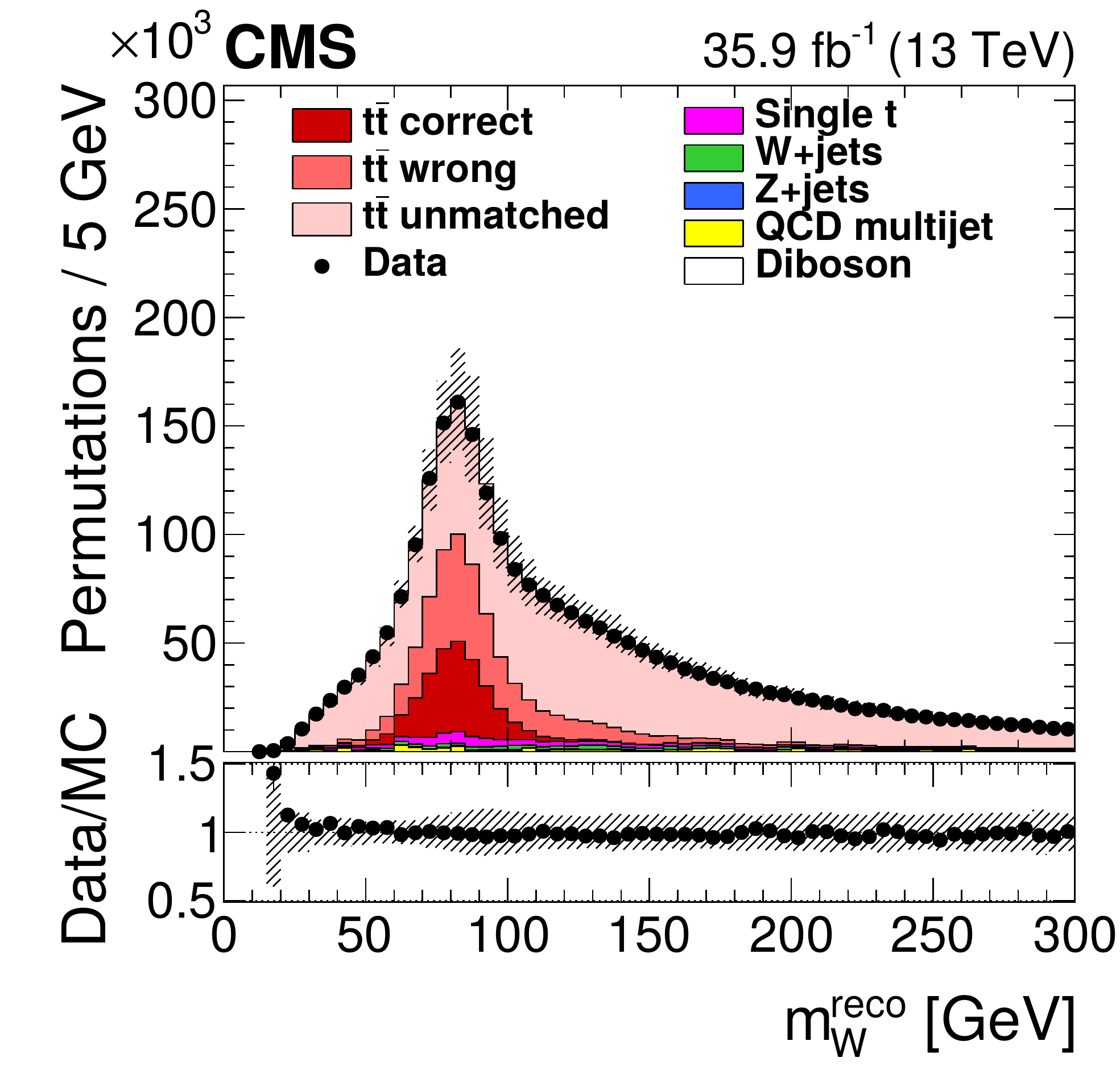}
\includegraphics[width=0.49\textwidth]{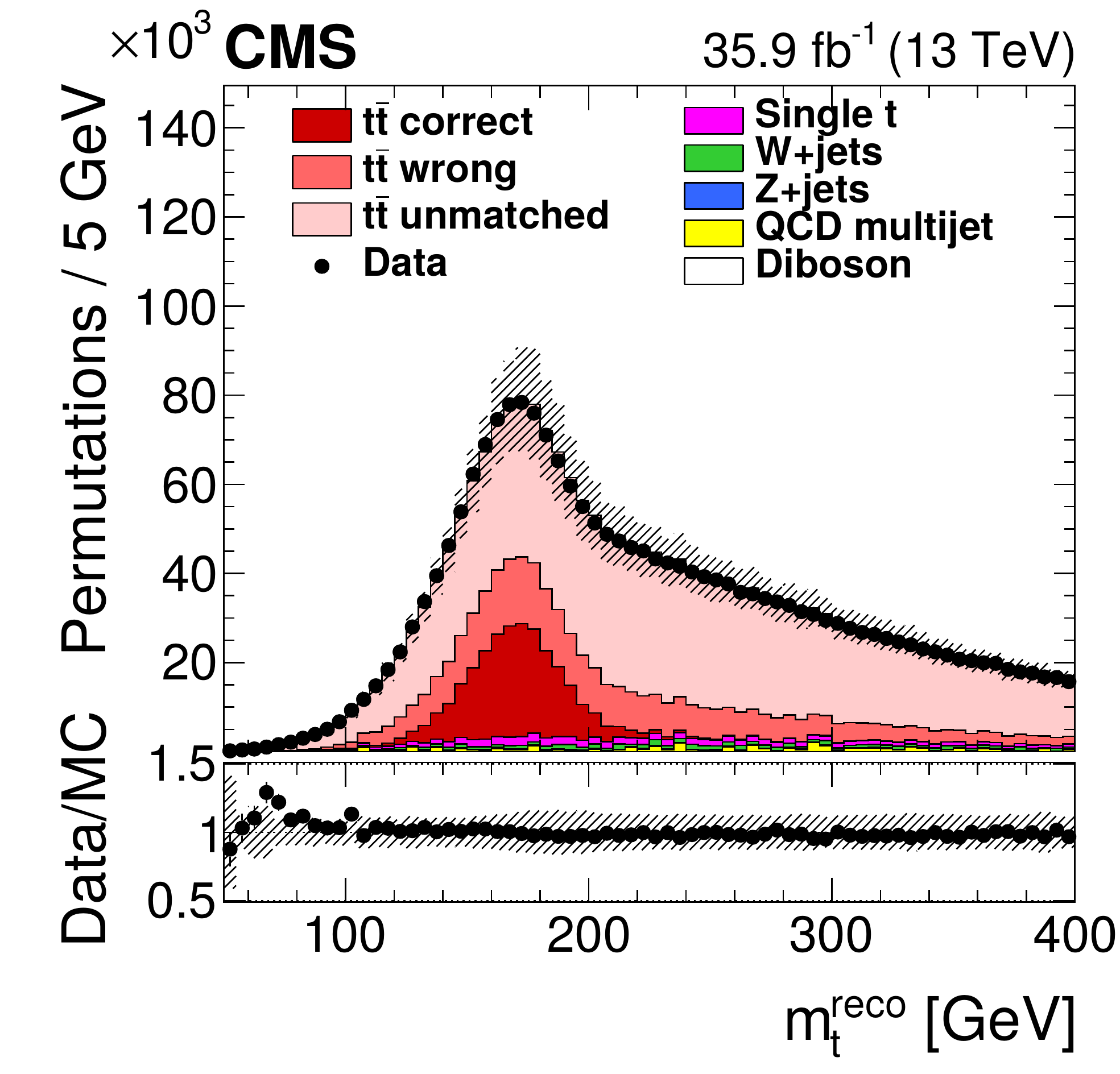}
\caption{\label{fig:controlplot-obs}
Invariant mass $m_\PW^\text{reco}$ of the two untagged jets (left) and invariant mass $\mtop^\text{reco}$ of the two untagged jets and one of the \cPqb-tagged jets (right) after the \cPqb~tagging requirement.
For the simulated \ttbar events, the jet-parton assignments are classified as correct, wrong, and unmatched permutations as described in the text.
The vertical bars show the statistical uncertainty  on the data and the hatched bands show the systematic uncertainties considered in Section~\ref{sec:systematics}.
The lower portion of each panel shows the ratio of the yields between data and the simulation.
The simulations are normalized to the integrated luminosity.}
\end{figure*}

In addition, at least four jets with $\pt>30\GeV$ and  $\abs{\eta} <2.4$ are required.
Only the four leading among the jets passing these $\pt$- and $\eta$-criteria are used in the reconstruction of the \ttbar system. 
Jets originating from \cPqb{} quarks are identified (tagged) using an algorithm that combines reconstructed secondary vertices and track-based lifetime information.
This has an efficiency of approximately 70\% and a mistagging probability for light-quark and gluon jets of 1\%~\cite{Sirunyan:2017ezt}.
We require exactly two \cPqb-tagged jets among the four leading ones and select  669\,109 \ttbar candidate events in data.
Figure~\ref{fig:controlplot-obs} shows the distributions of the reconstructed mass $m_\PW^\text{reco}$ of the \PW{} boson decaying to a  \qqbarpr{}  pair and the masses $\mtop^\text{reco}$ computed from the two untagged jets and each of the two \cPqb-tagged jets at this selection step.
For simulated \ttbar events, the parton-jet assignments can be classified as correct permutations (\emph{cp}), wrong permutations (\emph{wp}), and unmatched permutations (\emph{un}), where, in the latter, at least one quark from the \ttbar decay is not unambiguously matched within a distance of $\Delta R <0.4$ to any of the four selected jets.

{\tolerance 400
To check the compatibility of an event with the \ttbar hypothesis, and to improve the resolution of the reconstructed quantities, a kinematic fit~\cite{Abbott:1998dc} is performed.
For each event, the inputs to the algorithm are the four-momenta of the lepton and of the four leading jets, \ptvecmiss, and the resolutions of these variables.
The fit constrains these quantities to the hypothesis that two heavy particles of equal mass are produced, each one decaying to a bottom quark and a \PW{} boson, with the invariant mass of the latter constrained to 80.4\GeV.
The kinematic fit then minimizes $\chi^{2} \equiv \left(\mathbf{x}-\mathbf{x}^{m}\right)^\mathrm{T}G\left(\mathbf{x}-\mathbf{x}^{m}\right)$
where $\mathbf{x}^{m}$ and  $\mathbf{x}$ are the vectors of the measured and fitted momenta, respectively, and $G$ is the inverse covariance matrix which is constructed from the uncertainties in the measured momenta.
The two \cPqb-tagged jets are candidates for the \PQb{} quarks in the \ttbar hypothesis, while the two untagged jets serve as candidates for the light quarks from the hadronically decaying \PW{} boson.
This leads to two possible parton-jet assignments with two solutions for the longitudinal component of the neutrino momentum each, resulting in four different permutations per event.
\par}

\begin{figure*}[!tb]
\includegraphics[width=0.49\textwidth]{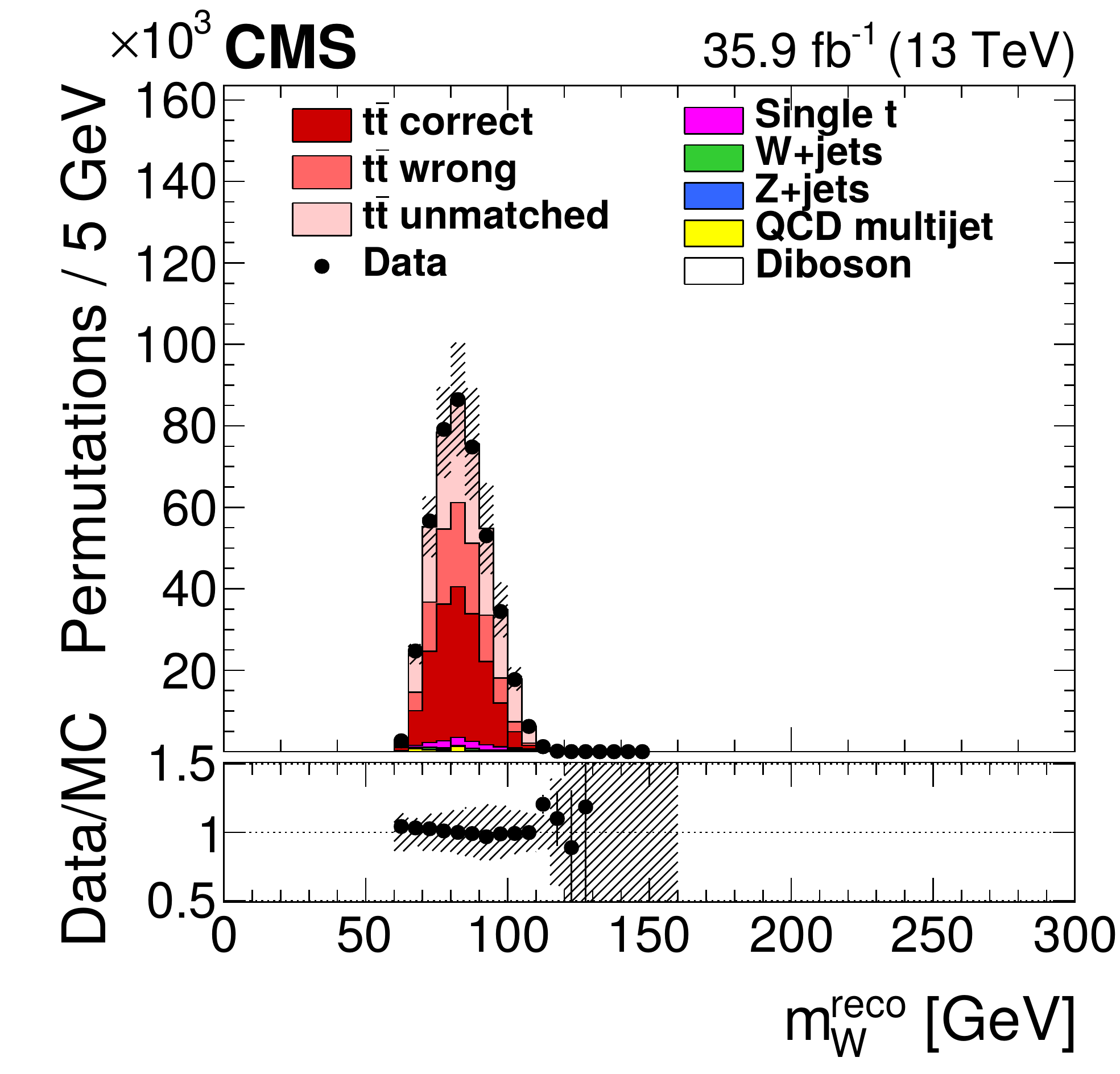}
\includegraphics[width=0.49\textwidth]{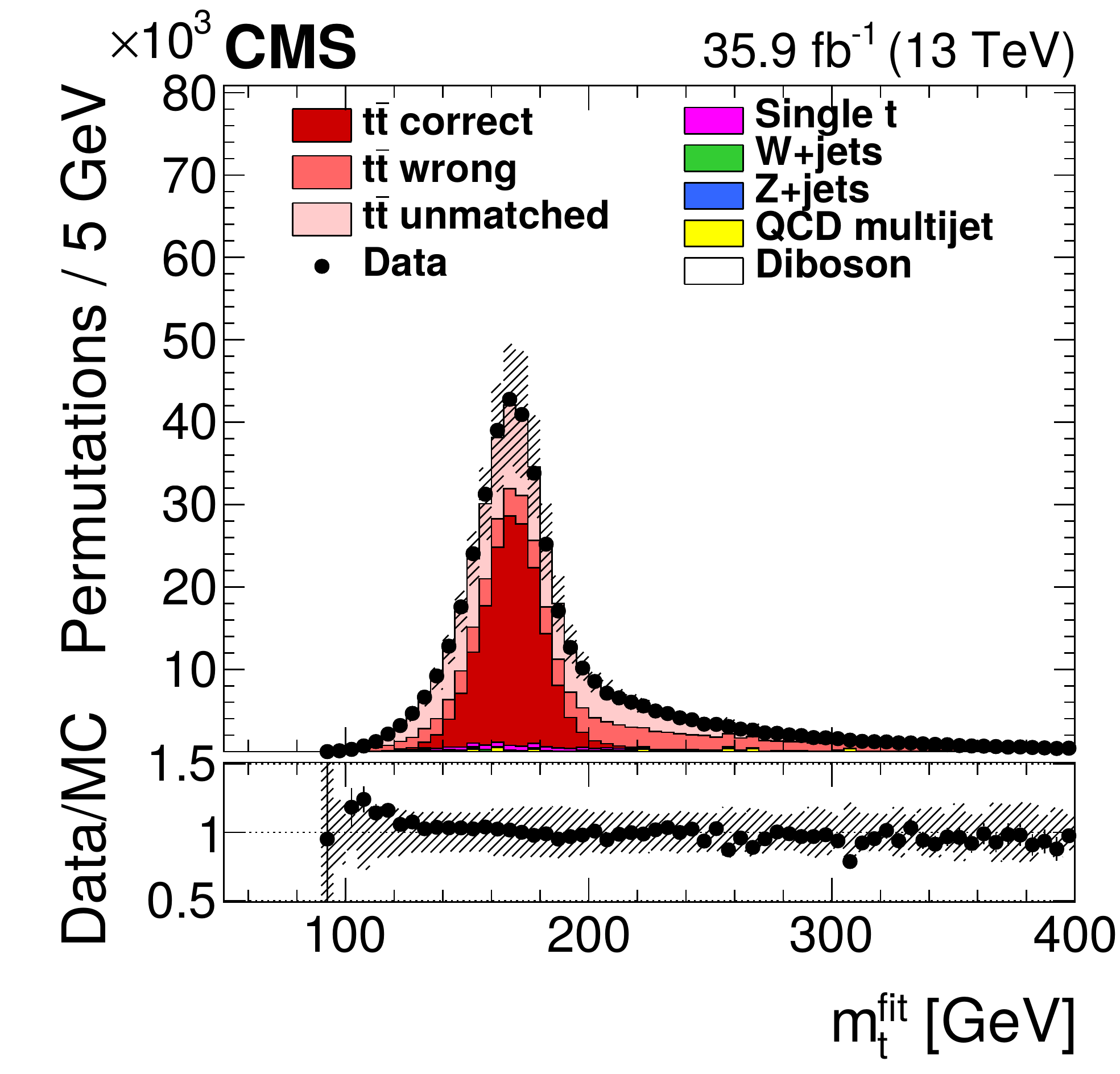}
\caption{\label{fig:controlplot-weighted-obs}
Reconstructed \PW{} boson masses $m_\PW^\text{reco}$ (left) and fitted top quark masses $\mtop^\text{fit}$ (right) after the goodness-of-fit selection and the weighting by $P_\text{gof}$.
Symbols and patterns are the same as in Fig.~\ref{fig:controlplot-obs}.
The simulations are normalized to the integrated luminosity.}
\end{figure*}

To increase the fraction of correct permutations, we require the goodness-of-fit (gof) probability for the kinematic fit with two degrees of freedom $P_\text{gof} = \exp\left(-\chi^{2}/2\right)$ to be at least $0.2$. This requirement selects 161\,496 events in data, while the non-\ttbar background in the simulated data is reduced from 7.6\% to 4.3\%. The remaining background consists mostly of single top quark events (2.5\%).
Any of the four permutations in an event that passes the  selection criteria is weighted by its $P_\text{gof}$ value and is used in the measurement.
These steps improve the fraction of correct permutations from 14.9\% to 48.0\%.
Figure~\ref{fig:controlplot-weighted-obs} shows the final distributions after the $P_\text{gof}$ selection of the reconstructed mass $m_\PW^\text{reco}$ of the \PW{} boson decaying to a \qqbarpr{} pair and the invariant mass of the top quark candidates from the kinematic fit $\mtop^\text{fit}$ for all selected permutations. These two observables are used in the mass extraction.

\section{Ideogram method}
\label{sec:ideogram}
An ideogram method~\cite{Abdallah:2008xh} is employed as described in Ref.~\cite{Chatrchyan:2012cz}.
The details of the procedure outlined below are identical with the approach taken in the Run~1 CMS measurement~\cite{Khachatryan:2015hba}.
The observable used to measure $\mtop$ is the mass $\mtop^\text{fit}$ evaluated after applying the kinematic fit.
We take the reconstructed \PW{} boson mass $m_\PW^\text{reco}$, before it is constrained by the kinematic fit, as an estimator for measuring the JSF to be applied in addition to the standard CMS JECs.
The top quark mass and the JSF are determined simultaneously in a likelihood fit to the selected permutations, in order to reduce the uncertainty from the JECs.

{\tolerance 1000
The distributions of $\mtop^\text{fit}$ and  $m_\PW^\text{reco}$ are obtained from simulation for seven different $\mtop$ and five different JSF values.
From these distributions, probability density functions $P_{j}$ are derived separately for the different permutation cases $j$: \emph{cp}, \emph{wp}, or \emph{un}. 
These functions depend on  $\mtop$ and the JSF and are labeled $P_{j}(m_{\cPqt,i}^\text{fit}|\mtop,\text{JSF})$ and  $P_{j}(m_{\PW, i}^\text{reco}|\mtop,\text{JSF})$, respectively, for the $i$th permutation of an event in the final likelihood.
The observables  $\mtop^\text{fit}$ and  $m_\PW^\text{reco}$ have a correlation coefficient with a size below 5\% for each permutation case and are treated as uncorrelated.
The most likely $\mtop$ and JSF values are obtained by minimizing $-2\ln \left[ \mathcal{L}\left(\text{sample} | \mtop,\text{JSF} \right) \right]$.
With an additional prior  $P(\text{JSF})$, the likelihood $\mathcal{L}\left(\text{sample} | \mtop,\text{JSF}\right)$ is defined as:
\begin{linenomath}
\ifthenelse{\boolean{cms@external}}
{ 
\begin{multline*}
  \mathcal{L}\left(\text{sample} | \mtop,\text{JSF} \right) =  P(\text{JSF}) \,\prod_\text{events}\bigg(\sum_{i=1}^{n} P_\text{gof}\left(i\right)\\
 \Big[\sum_{j}f_{j} \,P_{j}(m_{\cPqt,i}^\text{fit}|\mtop,\text{JSF}) \,P_{j}(m_{\PW ,i}^\text{reco}|\mtop,\text{JSF})\Big]\bigg)^{w_\text{evt}},
\end{multline*}
}
{ 
\begin{equation*}
 \mathcal{L}\left(\text{sample} | \mtop,\text{JSF} \right) =  P(\text{JSF}) \,\prod_\text{events}\left(\sum_{i=1}^{n} P_\text{gof}\left(i\right)\Big[\sum_{j}f_{j} \,P_{j}(m_{\cPqt,i}^\text{fit}|\mtop,\text{JSF})\, P_{j}(m_{\PW ,i}^\text{reco}|\mtop,\text{JSF})\Big]\right)^{w_\text{evt}},
\end{equation*}
}
\end{linenomath}
where  $n$ denotes the number of the at-most four permutations in each event, $j$ labels the permutation cases, and $f_j$ represents their relative fractions.
The event weight $w_\text{evt}=c\,\sum_{i=1}^{n}P_\text{gof}\left(i\right)$ is introduced to reduce the impact of events without correct permutations, where $c$ normalizes the average $w_\text{evt}$ to 1.
\par}

Different choices are made for the prior $P(\text{JSF})$ in the likelihood fit.
When the JSF is fixed to unity, the $P_{j}(m_{\PW ,i}^\text{reco}|\mtop,\text{JSF})$ can be approximated by a constant as they hardly depend on \mtop. Hence, only the $\mtop^\text{fit}$ observable is fit, and this approach is called the 1D analysis. The approach with an unconstrained JSF is called the 2D analysis.
Finally, in the hybrid analysis, the prior $P(\text{JSF})$ is a Gaussian centered at 1.0.
Its width depends on the relative weight $w_\text{hyb}$ that is assigned to the prior knowledge on the JSF, $\sigma_{\text{prior}}  = \delta\text{JSF}^{\text{2D}}_{\text{stat}} \sqrt{\smash[b]{1/w_\text{hyb} - 1}}$, where  $\delta\text{JSF}^{\text{2D}}_{\text{stat}}$ is the statistical uncertainty in the 2D result of the JSF.
The optimal value of  $w_\text{hyb}$ is determined  from the uncertainties in the 2D analysis and discussed in Section~\ref{sec:systematics}.

\label{sec:calibration}
{\tolerance 1800
The 2D method is separately calibrated for the muon and electron channel by conducting 10\,000 pseudo-experiments for each combination of the seven top quark masses  and the five $\text{JSF}$ values, using simulated \ttbar and background events.
We correct for deviations between the extracted mass and JSF and their input values. This bias correction amounts for the mass to an offset of 0.5\GeV for an expected value of 172.5\GeV, with a slope of 3\%.
Corrections for the statistical uncertainty of the method are derived from the widths of the corresponding pull distributions and have a size of 5\% for both the mass and the $\text{JSF}$.
\par}
\section{Systematic uncertainties}
\label{sec:systematics}
The systematic uncertainties in the  final measurement  are determined from pseudo-experiments.
Taking into account new simulations, more variations of the modeling of the \ttbar events are investigated than in the Run~1 analysis~\cite{Khachatryan:2015hba}. The scales used for the simulation of initial-state radiation (ISR) and final-state radiation (FSR) are varied independently from the renormalization and factorization scales.
Furthermore, the effects of early resonance decays and alternative color-reconnection models~\cite{Argyropoulos:2014zoa,Christiansen:2015yqa} are evaluated, while in Run~1 only the effect of an underlying event tune without color reconnection was studied.
The relevant systematic uncertainties and the methods used to evaluate them are described below.
{\tolerance 8000
\begin{description}
\item [{{Method calibration:}}] We consider the quadratic sum of statistical uncertainty and residual biases after the calibration of the ideogram method as a systematic uncertainty.
\item [{{JECs:}}] As we measure a global JSF, we have to take into account the influence of the \pt- and $\eta$-dependent JEC uncertainties.
This is done by scaling the energies of all jets up and down according to their individual uncertainties~\cite{Khachatryan:2016kdb}, split into correlation groups (called InterCalibration, MPFInSitu and Uncorrelated) similarly to the procedure adopted at 8\TeV~\cite{CMS-PAS-JME-15-001}.
\item [{{Jet energy resolution:}}] The jet energy resolution (JER) in simulation
is slightly degraded to match the resolutions measured in data~\cite{Khachatryan:2016kdb}. To account for the resolution uncertainty, the JER in the simulation is modified by ${\pm}1$ standard deviation with respect to the degraded resolution.
\item [{{\ifthenelse{\boolean{cms@external}}{$\PQb{}$}{$\boldsymbol{\PQb{}}$} tagging:}}] The events are weighted to account for the \pt-dependent uncertainty of the \PQb{} tagging efficiencies and misidentification rates of the \PQb{} tagging algorithm~\cite{Sirunyan:2017ezt}.
\item [{{Pileup:}}] To estimate the uncertainties associated with the determination of the number of pileup events and with the weighting procedure, the inelastic \Pp\Pp{} cross section is varied by ${\pm}4.6$\% for all simulations.
\item [{{Non-\ifthenelse{\boolean{cms@external}}{\ttbar}{$\boldsymbol{\ttbar}$} background:}}] The main uncertainty in the non-\ttbar background stems from the uncertainty in the measurements of the cross sections used in the normalization. The normalization of the background samples is varied by $\pm$10\% for the single top quark samples~\cite{Khachatryan:2014iya,Sirunyan:2016cdg}, $\pm$30\% for the \PW{}+jets samples~\cite{Khachatryan:2016ipq}, $\pm$10\% for the \PZ{}+jets~\cite{Khachatryan:2016iob} and for the diboson samples~\cite{Khachatryan:2016tgp,Sirunyan:2017zjc}, and $\pm$100\% for the QCD multijet samples.
The uncertainty in the luminosity of 2.5\%~\cite{CMS-PAS-LUM-17-001} is negligible compared to these variations.
\item [{{JEC Flavor:}}] The Lund string fragmentation implemented in \PYTHIA~6.422~\cite{Sjostrand:2006za} is compared to the cluster fragmentation of \HERWIGpp~2.4~\cite{Bahr:2008pv}.
Each model relies on a large set of tuning parameters that allow to modify the individual fragmentation of jets initiated from gluons, light quarks, and \PQb{} quarks.
Therefore, the difference in jet energy response between \PYTHIA6  and \HERWIGpp is determined for each jet flavor~\cite{Khachatryan:2016kdb}.
In order to evaluate possible differences between the measured JSF (from light quarks with gluon contamination) and the \PQb{} jet energy scale, the flavor uncertainties for jets from light quarks, gluons, and bottom quarks are evaluated separately and added linearly.
\item [{{\ifthenelse{\boolean{cms@external}}{$\PQb{}$}{$\boldsymbol{\PQb{}}$} jet modeling:}}] This term has three components:
The fragmentation into \PQb{} hadrons is varied in simulation within the uncertainties of the Bowler--Lund fragmentation function tuned to  ALEPH~\cite{Heister:2001jg} and DELPHI~\cite{DELPHI:2011aa} data. In addition, the difference between the Bowler--Lund~\cite{Bowler:1981sb} and the Peterson~\cite{Peterson:1982ak} fragmentation functions is included in the uncertainty.
Lastly, the uncertainty from the semileptonic \PQb{} hadron branching fraction is obtained by varying it by $-0.45$\% and $+0.77$\%, which is the range of the measurements from \PBz/\PBp{} decays and their uncertainties~\cite{Olive:2016xmw}.
\item [{{PDFs:}}] The NNPDF3.0 NLO ($\alpS=0.118$) PDF is used in the generation of simulated events. We calculate the results with the different PDF replicas and use the variance of these predictions for the PDF uncertainty~\cite{Ball:2014uwa}. In addition, NNPDF3.0 sets with $\alpS = 0.117$ and 0.119 are evaluated and the observed difference is added in quadrature~\cite{Rojo:2015acz, Butterworth:2015oua, Accardi:2016ndt}.
\item [{{Renormalization and factorization scales:}}] The simulated events are weighted to match the event shape distributions generated with different renormalization and factorization scales. These scales are varied independently from each other by a factor of 0.5 and 2.
\item [{{ME/PS matching:}}] The model parameter $h_{\text{damp}}=1.58^{+0.66}_{-0.59}$~\cite{CMS-PAS-TOP-16-021}  used in \POWHEG to control the matching of the MEs to the \PYTHIA~8 PS is varied within its uncertainties.
\item [{{ME generator:}}]
The influence of the NLO ME generator and its matching to the PS generator is estimated by using a sample from the NLO generator \MGvATNLO with \textsc{FxFx} matching~\cite{Frederix:2012ps}, instead of the \POWHEG~v2 generator used as default.
\item [{{ISR PS scale:}}]
The PS scale value used for the simulation of ISR in \PYTHIA~8 is scaled up by 2 and down by 0.5 in dedicated samples.
\item [{{FSR PS scale:}}]
The PS scale value used for the simulation of FSR in \PYTHIA~8 is scaled up by $\sqrt{2}$ and down by $1/\sqrt{2}$~\cite{Skands:2014pea} in dedicated samples.
This affects the fragmentation and hadronization of the jets initiated by the ME calculation, as well as the emission of extra jets.
In the FSR samples, the jet energy response of the light quarks is observed to differ by $\pm 1.2$\% compared to the response of the default sample. This response difference would be absorbed in the residual JECs if the  corrections were derived based on $\gamma$/\PZ{}+jet simulations with the same PS scale. Hence, the momenta of all jets in the varied samples are scaled so that the energy response for jets induced by light quarks agrees with the default sample.
\item [{{Top quark  \ifthenelse{\boolean{cms@external}}{$\pt$}{$\boldsymbol{\pt}$}:}}] Recent calculations~\cite{Czakon:2015owf} suggest that next-to-next-to-leading-order effects have an important impact on the top quark \pt  spectrum, that NLO ME generators are unable to reproduce. Therefore, the top quark \pt in simulation is varied to match the distribution measured by CMS~\cite{Khachatryan:2016mnb, Sirunyan:2017mzl}. The observed difference with respect to the default sample is quoted as a systematic uncertainty.
\item [{{Underlying event:}}] The modeling of multiple-parton interactions in \PYTHIA~8 is tuned to measurements of the underlying event~\cite{Skands:2014pea, CMS-PAS-TOP-16-021}. The parameters of the tune are varied within their uncertainties in the simulation of the \ttbar signal.
\item [{{Early resonance decays:}}]
By enabling early resonance decays (ERDs) in \PYTHIA~8,   color reconnections can happen between particles from the  top quark decay and particles from the underlying event. In the default sample the ERDs are turned off and the top quark decay products do not interact with the underlying event.
The influence of the ERD setting is estimated from a sample with ERDs enabled in \PYTHIA~8.
\item [{{Color reconnection:}}]
The uncertainties that arise from ambiguities in modeling color-re\-con\-nec\-tion effects are estimated by comparing the default model in \PYTHIA~8 with ERDs to two alternative models of color reconnection, a model with string formation beyond leading color (``QCD inspired'')~\cite{Christiansen:2015yqa} and a model that allows gluons to be moved to another string (``gluon move'')~\cite{Argyropoulos:2014zoa}. All models are tuned to measurements of the underlying event~\cite{Skands:2014pea, Sirunyan:2018avv}.
The observed shifts are listed in Table~\ref{tab:cr}.
\begin{table*}[!htb]
\topcaption{\label{tab:cr} Observed shifts with respect to the default simulation for different models of color reconnection. The ``QCD inspired'' and ``gluon move'' models are compared to the default model with ERDs.  The statistical uncertainty in the JSF shifts is  0.1\%.}
\centering{}
\begin{tabular}{lccccc}
 & \multicolumn{2}{c}{2D approach} & 1D approach & \multicolumn{2}{c}{Hybrid}\\
 & $\delta m_{\PQt}^{\text{2D}}$ & $\delta\text{JSF}^{\text{2D}}$  & $\delta m_{\PQt}^{\text{1D}}$ & $\delta m_{\PQt}^{\text{hyb}}$ & $\delta\text{JSF}^{\text{hyb}}$ \\
  & [\GeVns{}] & [\%] & [\GeVns{}] & [\GeVns{}] & [\%] \\
\hline
\POWHEG \textsc{p8} ERD on  & $-0.22\pm0.09$ & $+0.8$ & $+0.42\pm0.05$ & $-0.03\pm0.07$ & $+0.5$\\
\POWHEG \textsc{p8} QCD inspired & $-0.11\pm0.09$ & $-0.1$ & $-0.19\pm0.06$ & $-0.13\pm0.08$ & $-0.1$ \\
\POWHEG \textsc{p8} gluon move & $+0.34\pm0.09$ & $-0.1$ & $+0.23\pm0.06$ & $+0.31\pm0.08$ & $-0.1$ \\
\end{tabular}
\end{table*}
Among the two approaches, the ``gluon move'' model leads to larger shifts and these are quoted as the systematic uncertainty.
\end{description}
\par}

The modeling uncertainties are mainly evaluated by varying the parameters within one model: \POWHEG~v2 + \PYTHIA~8 with the \textsc{CUETP8M2T4} tune (labeled as \POWHEG \textsc{p8} M2T4).
This approach benefits from the calibration of the reconstructed physics objects which is derived from data with \PYTHIA~8 as a reference.
Three alternative models of the \ttbar signal are studied. The NLO \MGvATNLO generator with  the {FxFx} matching~\cite{Frederix:2012ps} (labeled as MG5 \textsc{p8} [FxFx] M2T4) and the LO  \MGvATNLO with the {MLM} matching~\cite{Alwall:2007fs} (labeled as MG5 \textsc{p8} [MLM] M1) are both interfaced with \PYTHIA~8 with the \textsc{CUETP8M2T4} and the {CUETP8M1} tune, respectively.
In addition, \POWHEG~v2 interfaced with \HERWIGpp~\cite{Bahr:2008pv} (v2.7.1) with the tune {EE5C}~\cite{Seymour:2013qka} (labeled as \POWHEG \textsc{h++} EE5C) is evaluated. ME corrections to the top quark decay are not applied in the \HERWIGpp sample.
A dedicated analysis has found that MG5 \textsc{p8} [MLM] M1 and \POWHEG \textsc{h++} EE5C do not describe the data well~\cite{CMS-PAS-TOP-16-021, Sirunyan:2018wem} and only the NLO MG5 \textsc{p8} [FxFx] M2T4 model is used in the evaluation of the systematic uncertainties.

Nevertheless, the analysis is also performed on pseudo-experiments where the \ttbar signal stems from these different generator setups.
This yields rather large shifts for the two discarded models. The results are summarized in Table~\ref{tab:gen}.
The shift for \POWHEG \textsc{h++} EE5C would translate into a 4\GeV higher measurement of $\mtop$ if this setup were used as the default \ttbar simulation and not as signal in the pseudo-data.
The agreement of these generator setups and the color-reconnection models with data are studied in Section~\ref{sec:diff} for this top quark mass measurement.
\begin{table*}[!htb]
\topcaption{\label{tab:gen} Observed shifts with respect to the default simulation for different generator setups.  The statistical uncertainty in the JSF shifts is  0.1\%.}
\centering
\begin{tabular}{lccccc}
 & \multicolumn{2}{c}{2D approach} & 1D approach & \multicolumn{2}{c}{Hybrid}\\
 & $\delta m_{t}^{\text{2D}}$ & $\delta\text{JSF}^{\text{2D}}$  & $\delta m_{t}^{\text{1D}}$   & $\delta m_{t}^{\text{hyb}}$ & $\delta\text{JSF}^{\text{hyb}}$ \\
 & [\GeVns{}] & [\%] & [\GeVns{}] & [\GeVns{}] & [\%] \\
\hline
MG5 \textsc{p8} [FxFx] M2T4 &  $+0.15\pm0.23$  & $+0.2$  & $+0.32\pm0.14$  & $+0.20\pm0.19$  & $+0.1$\\
MG5 \textsc{p8} [MLM] M1 &  $+0.82\pm0.16$ & $<$0.1 & $+0.80\pm0.10$& $+0.82\pm0.14$ & $<$0.1  \\
\POWHEG \textsc{h++} EE5C &  $-4.39\pm0.09$ & $+1.4$ & $-3.26\pm0.06$ & $-4.06\pm0.08$ & $+1.0$ \\
\end{tabular}
\end{table*}

The contributions from the different sources of systematic uncertainties are shown in Table~\ref{tab:Systematic-uncertainties}.
In general, the absolute value of the largest observed shifts in $\mtop$ and JSF, determined  by changing the parameters by $\pm$1 standard deviation ($\sigma$), are assigned as systematic uncertainties.
The only exception to this is if the statistical uncertainty in the observed shift is larger than the value of the calculated shift. In this case the statistical uncertainty is taken as the best estimate of the uncertainty in the parameter.
The signs in the table are taken from the $+1\sigma$ shift in the value of the uncertainty source where applicable.

\begin{table*}[!htb]
\topcaption{\label{tab:Systematic-uncertainties}List of systematic uncertainties for the fits to the combined data set using the procedures described in Section~\ref{sec:systematics}.
With the exception of the flavor-dependent JEC terms, the total systematic uncertainty is obtained from the sum in quadrature of the individual systematic uncertainties.
 The values in parentheses with indented labels are already included in the preceding uncertainty source.
A positive sign indicates an increase in the value of \mtop or the JSF in response to a $+1\sigma$ shift and a negative sign indicates a decrease.
The statistical uncertainty in the shift in $\mtop$ is given when different samples are compared. The statistical uncertainty in the JSF shifts is  0.1\% for these sources.
}
\centering
\begin{tabular}{lccccc}
 & \multicolumn{2}{c}{2D approach} & 1D approach & \multicolumn{2}{c}{Hybrid}\\
 & $\delta m_{t}^{\text{2D}}$ & $\delta\text{JSF}^{\text{2D}}$  & $\delta m_{t}^{\text{1D}}$ & $\delta m_{t}^{\text{hyb}}$ & $\delta\text{JSF}^{\text{hyb}}$ \\
  & [\GeVns{}] & [\%] & [\GeVns{}] & [\GeVns{}] & [\%] \\
\hline
\textit{Experimental uncertainties} &  \multicolumn{5}{c}{}\\
Method calibration  & 0.05  & $<$0.1  & 0.05  & 0.05 & $<$0.1\\
JEC (quad. sum) &0.13& 0.2 & 0.83 & 0.18 & 0.3\\
-- InterCalibration  & ($-0.02$) & ($<$0.1)  & ($+0.16$)  & ($+0.04$) & ($<$0.1) \\
-- MPFInSitu  & ($-0.01$) & ($<$0.1)  & ($+0.23$) & ($+0.07$) & ($<$0.1) \\
-- Uncorrelated  & ($-0.13$) & ($+0.2$) & ($+0.78$) & ($+0.16$) & ($+0.3$)\\
Jet energy resolution  & $-0.20$ & $+0.3$ & $+0.09$ & $-0.12$ & $+0.2$\\
\PQb{} tagging  & $+0.03$ & $<$0.1 & $+0.01$ & $+0.03$ & $<$0.1 \\
Pileup  & $-0.08$ & $+0.1$ & $+0.02$ & $-0.05$ & $+0.1$\\
Non-\ttbar background  & $+0.04$ &  $-0.1$ & $-0.02$ & $+0.02$ & $-0.1$ \\ [\cmsTabSkip]
\textit{Modeling uncertainties} &  \multicolumn{5}{c}{}\\
JEC Flavor (linear sum) & $-0.42$ & $+0.1$ & $-0.31$ & $-0.39$ & $<$0.1\\
-- light quarks (uds) & ($+0.10$) & ($-0.1$) & ($-0.01$) & ($+0.06$) & ($-0.1$) \\
-- charm & ($+0.02$) &  ($<$0.1) & ($-0.01$) & ($+0.01$) & ($<$0.1) \\
-- bottom & ($-0.32$) &   ($<$0.1) & ($-0.31$) & ($-0.32$) &  ($<$0.1) \\
-- gluon & ($-0.22$) & ($+0.3$) & ($+0.02$) & ($-0.15$) & ($+0.2$) \\
\PQb jet modeling (quad. sum)  & 0.13 & 0.1 & 0.09 & 0.12 & $<$0.1 \\
-- \PQb{} frag. Bowler--Lund & ($-0.07$) & ($+0.1$) & ($-0.01$) & ($-0.05$) & ($<$0.1)\\
-- \PQb{} frag. Peterson  & ($+0.04$) & ($<$0.1) & ($+0.05$) & ($+0.04$) & ($<$0.1)\\
-- semileptonic \PB{} decays  & ($+0.11$) & ($<$0.1) & ($+0.08$) & ($+0.10$) & ($<$0.1) \\
PDF  & 0.02 & $<$0.1 & 0.02 & 0.02 & $<$0.1  \\
Ren. and fact. scales & 0.02 & 0.1 & 0.02 & 0.01 & $<$0.1\\
ME/PS matching   & $-0.08\pm0.09$ & $+0.1$ & $+0.03\pm0.05$ & $-0.05\pm0.07$ & $+0.1$\\
ME generator  & $+0.15\pm0.23$  & $+0.2$  & $+0.32\pm0.14$  & $+0.20\pm0.19$  & $+0.1$\\
ISR PS scale & $+0.07\pm0.09$ & $+0.1$ & $+0.10\pm0.05$ & $+0.06\pm0.07$ & $<$0.1 \\
FSR PS scale & $+0.24\pm0.06$ & $-0.4$ & $-0.22\pm0.04$ & $+0.13\pm0.05$ & $-0.3$ \\
Top quark \pt & $+0.02$ & $-0.1$ & $-0.06$ & $-0.01$ & $-0.1$\\
Underlying event  & $-0.10 \pm 0.08$ & $+0.1$ & $+0.01\pm 0.05$ & $-0.07\pm 0.07$ & $+0.1$ \\
Early resonance decays & $-0.22\pm0.09$ & $+0.8$ & $+0.42\pm0.05$ & $-0.03\pm0.07$ & $+0.5$ \\
Color reconnection & $+0.34\pm0.09$ & $-0.1$ & $+0.23\pm0.06$ & $+0.31\pm0.08$ & $-0.1$ \\ [\cmsTabSkip]
\textbf{Total systematic} &  \textbf{0.75} & \textbf{1.1} & \textbf{1.10} & \textbf{0.62 } & \textbf{0.8}\\
Statistical (expected) & 0.09 & 0.1 & 0.06 & 0.08 & 0.1 \\  [\cmsTabSkip]
\textbf{Total (expected)} & \textbf{0.76} & \textbf{1.1} & \textbf{1.10} & \textbf{0.63} & \textbf{0.8}\\
\end{tabular}
\end{table*}

The details of the fitting procedure have several consequences on the uncertainties.
The inclusion of the JSF as a nuisance parameter in the fit and its constraint by the $m_\PW^\text{reco}$ observable reduces not only the uncertainties stemming from the JECs, but also the modeling uncertainties.
As the JSF is an overall energy scale factor derived mainly on light-quark jets and applied to all jets, this approach cannot reduce the uncertainties on the flavor-dependent JECs.
The other remaining systematic uncertainties are also dominated by effects that cannot be fully compensated through the simultaneous determination of $\mtop$ and JSF, \ie, the $\mtop^\text{fit}$ observable is affected differently from $m_\PW^\text{reco}$.
For the hybrid analysis, a hybrid weight of  $w_\text{hyb}=0.3$ is found optimal based on the total uncertainty in the 2D result of the JSF and the jet energy scale uncertainty in the JECs.
Due to the larger jet energy uncertainties at the beginning of the 13\TeV data taking, $w_\text{hyb}$ is lower than in the Run~1 analysis~\cite{Khachatryan:2015hba} where the prior JSF knowledge contributes 50\% of the information.
With an expected statistical uncertainty $\delta\text{JSF}^{\text{2D}}_{\text{stat}} = 0.08\%$ on the JSF for the 2D analysis, the width of the prior is $\sigma_{\text{prior}}  = 0.12\%$.
The hybrid analysis leads to further reduced uncertainties in the FSR PS scale and in ERDs  compared to the 2D analysis.
This stems from the opposite signs of the observed shifts in \mtop for the 1D and 2D analyses, \ie, the JSF from the 2D analysis overcompensates the effects on $\mtop^\text{fit}$.

\section{Results}
\label{sec:results}
The 2D fit to the selected lepton+jets events yields:
\begin{equation*}
\begin{aligned}
\mtop^{\text{2D}} & =  172.40\pm0.09\statJSF\pm0.75\syst\GeV,\\
\mathrm{JSF}^{\text{2D}} & =  0.994\pm0.001\stat\pm0.011\syst.
\end{aligned}
\end{equation*}
As the top quark mass  and the JSF are measured simultaneously, the statistical uncertainty in $\mtop$ originates from both quantities of interest. The measured unconstrained JSF is compatible with the one obtained from jets recoiling against photons and \PZ{} bosons within its uncertainties.

Separate fits to the 101\,992 muon+jets events and the 59\,504 electron+jets events give statistically compatible results:
\ifthenelse{\boolean{cms@external}}
{ 
\begin{equation*}
\begin{aligned}
\mu\text{+jets: } & \mtop^{\text{2D}}  =  172.44\pm0.11\statJSF \GeV,\\
                                   & \text{JSF}^{\text{2D}}  = 0.995\pm 0.001\stat,\\
\Pe\text{+jets: } & \mtop^{\text{2D}}  =  172.32\pm0.16\statJSF \GeV,\\
                           & \text{JSF}^{\text{2D}}  = 0.993\pm0.001\stat.\\
\end{aligned}
\end{equation*}
}
{
\begin{equation*}
\begin{aligned}
\mu\text{+jets: } &\mtop^{\text{2D}} & =  172.44\pm0.11\statJSF\GeV,\ &\mathrm{JSF}^{\text{2D}} = 0.995\pm 0.001\stat,\\
\Pe\text{+jets: }   &\mtop^{\text{2D}} & =  172.32\pm0.16\statJSF\GeV,\ &\mathrm{JSF}^{\text{2D}} = 0.993\pm0.001\stat.\\
\end{aligned}
\end{equation*}
}
The 1D fit and the hybrid fit with $w_\text{hyb}=0.3$, as obtained in Section~\ref{sec:systematics},  yield for the lepton+jets channel:
\begin{equation*}
\begin{aligned}
\mtop^{\text{1D}} & =  171.93\pm0.06\stat\pm1.10\syst\GeV,\\
\mtop^{\text{hyb}} & =  172.25\pm0.08\statJSF\pm0.62\syst\GeV,\\
\mathrm{JSF}^{\text{hyb}} & =  0.996\pm0.001\stat\pm0.008\syst.
\end{aligned}
\end{equation*}

{\tolerance 1400
The hybrid fit measurement of $\mtop = 172.25\pm 0.08\statJSF\pm 0.62\syst\GeV$
offers the lowest overall uncertainty and, therefore, is chosen as the main result of this study.
This is the first published result of the top quark mass measured with Run~2 data and the new NLO generator setups.
Because of the larger integrated luminosity and the higher \ttbar cross section at $\sqrt{s}=13\TeV$, the statistical uncertainty is halved compared to the Run~1 result of $\mtop = 172.35\pm 0.16\statJSF\pm 0.48\syst$\GeV~\cite{Khachatryan:2015hba}.
This measurement is consistent with the Run~1 result within the uncertainties.
The previous measurement was calibrated with \ttbar events generated at LO with \MADGRAPH~5.1.5.11~\cite{Alwall:2011uj} matched to \PYTHIA~6.426 PS~\cite{Sjostrand:2006za} with the  Z2$^*$ tune~\cite{Field:2010bc} using the \textsc{MLM} prescription.
No shift in the measured top quark mass from the new simulation at NLO with \POWHEG~v2 and \PYTHIA~8 and the new experimental setup is observed.
The systematic uncertainties are larger than for the Run~1 result due to a more advanced treatment of the modeling uncertainties.
This is mainly caused by the evaluation of a broader set of color-reconnection models that were not available in Run~1, yielding a more extensive treatment of the associated uncertainty.
Without the uncertainty due to these models of $0.31\GeV$, the systematic uncertainties in \mtop would be reduced from 0.62 to 0.54\GeV and would be much closer to the Run~1 result.
Tighter constraints on the existing color-reconnection models and the settings in the NLO simulations can occur in the near future and reduce the systematic uncertainties due to these specific models.
The new treatment of the modeling uncertainties will require special care when combining this measurement with the Run~1 result.
\par}
\section{Measured top quark mass as a function of kinematic observables\label{sec:kine}}
\label{sec:diff}
The modeling of soft and perturbative QCD effects is the main source of systematic uncertainties on the analysis presented here.
Differential measurements of \mtop as a function of the kinematic properties of the \ttbar system can be used to validate the different models and to identify possible biases in the measurement.
Variables are selected that probe potential effects from color reconnection, ISR and FSR, and the kinematic observables of the jets coming from the top quark decays. 
They are the transverse momentum of the hadronically decaying top quark ($\pt^{\rm{t,had}}$), the invariant mass of the \ttbar system ($m_{\ttbar}$), the transverse momentum of the \ttbar system ($\pt^{\ttbar}$), the number of jets with $\pt > 30\GeV$ ($N_\text{jets}$), the \pt and the pseudorapidity of the \PQb{} jet assigned to the hadronic decay branch ($\pt^{\PQb,\text{had}}$ and $\abs{\eta^{\PQb,\text{had}}}$), the $\Delta R$ between the \PQb{} jets ($\Delta R_{\bbbar}$), and the $\Delta R$ between the light-quark jets ($\Delta R_{\qqbarpr}$).
These are the same variables  as in the Run~1 analysis~\cite{Khachatryan:2015hba}.

For each variable, the event sample is divided into three to five bins as a function of the value of this variable, and we populate each bin using all permutations which lie within the bin boundaries.
As some variables depend on the parton-jet assignment that cannot be resolved unambiguously, such as the \pt of a reconstructed top quark, a single event is allowed to contribute to multiple bins.
For each bin, \mtop is measured using the hybrid likelihood fit with the same probability density functions as for the inclusive measurement.  
The JSF prior is chosen such that it constrains the measured JSF with the same relative strength.
This procedure was also used in the Run~1 analysis~\cite{Khachatryan:2015hba}.

{\tolerance=800
For the modeling of the perturbative QCD effects, the data are compared to the MG5 \textsc{p8} [FxFx] M2T4,  MG5 \textsc{p8} [MLM] M1, and \POWHEG \textsc{h++} EE5C setups.
For the modeling of color reconnection, the default tune of \PYTHIA~8, the ``QCD inspired'' model~\cite{Christiansen:2015yqa}, and the ``gluon move'' model~\cite{Argyropoulos:2014zoa} are considered. The three latter models are simulated with ERDs in \PYTHIA~8.
\par}

In these comparisons, the mean value of the measured top quark mass is subtracted from the measurement in each bin of the sample and the results are expressed in the form of offsets $\mtop-\left<\mtop\right>$, where the mean comes from the inclusive measurement on the specific sample. The subtracted offsets with respect to \POWHEG \textsc{p8} M2T4 can be found in the Tables~\ref{tab:cr} and \ref{tab:gen}.
To aid in the interpretation of a difference between the value of $\mtop-\left<\mtop\right>$ and the prediction from a simulation in the same bin, a bin-by-bin calibration of the results is applied. 
This is derived using the \POWHEG \textsc{p8} M2T4 simulation with  the same technique as for the inclusive measurement except that it is performed for each bin separately. 
The bin-by-bin bias correction for the mass can be much larger than for the inclusive analysis and reaches up to 10\GeV for some bins.
For each bin the statistical uncertainty and the dominant systematic uncertainties are combined in quadrature, where the latter include JEC (\pt-, $\eta$-, and flavor-dependent), JER, pileup, \PQb{} fragmentation, renormalization and factorization scales, ME/PS matching, ISR/FSR PS scales, and the underlying event.

For each variable and model, the cumulative $\chi^2$ between the model and the data is computed taking into account the statistical uncertainty in the model prediction and the total uncertainty in the data value.
 The number of degrees of freedom for each variable is the number of bins minus one as the mean measured top quark mass is subtracted. The resulting $\chi^2$ probabilities ($p$-values) are listed in Table~\ref{tab:diff}.
\begin{table*}[!htb]
\topcaption{\label{tab:diff} Compatibility of different models with the differential measurement of the top quark mass. For each variable and model, the probability of the cumulative $\chi^2$ is computed.
The setup with \POWHEG~v2 + \HERWIGpp does not use ME corrections to the top quark decay and shows large deviations from the data.}
\noindent \centering{}
\setlength\tabcolsep{3.5pt}
\begin{tabular}{lcccccccc}
 \multirow{2}{*}{Model} & \multicolumn{8}{c}{$\chi^2$ probability} \\
\rule[-1em]{0pt}{1em} & $\pt^{\PQt,\text{had}}$ & $m_{\ttbar}$ & $\pt^{\ttbar}$ & $N_\text{jets}$ & $\pt^{\PQb,\text{had}}$ & $\abs{\eta^{ \PQb,\text{had}}}$ & $\Delta R_{\bbbar}$ & $\Delta R_{\qqbarpr}$ \\
\hline
\POWHEG \textsc{p8} M2T4                  & 0.68 & 0.94 & 0.91 & 0.71 & 0.98 & 0.60 & 0.61 & 0.70 \\
MG5 \textsc{p8} [FxFx] M2T4               & 0.98 & 0.78 & 0.93 & 0.94 & 0.80 & 0.35 & 0.94 & 0.91 \\
MG5 \textsc{p8} [MLM] M1                   & 0.48 & 0.84 & 0.99 & 0.41 & 0.98 & 0.17 & 0.71 & 0.61 \\
\POWHEG \textsc{h++} EE5C              & 0.07 & $2{\times}10^{-13}$ & 0.52 & 0.72 & $2{\times}10^{-4}$ & 0.55 & 0.36 & $2{\times}10^{-5}$ \\
\POWHEG \textsc{p8} ERD on              & 0.75 & 0.99 & 0.83 & 0.53 & 0.95 & 0.64 & 0.38 & 0.96 \\
\POWHEG \textsc{p8} QCD inspired    & 0.80 & 0.94 & 0.94 & 0.66 & 0.99 & 0.71 & 0.49 & 0.90 \\
\POWHEG \textsc{p8} gluon move      & 0.87 & 0.94 & 0.93 & 0.72 & 0.93 & 0.51 & 0.59 & 0.93 \\
\end{tabular}
\end{table*}

\begin{figure*}[!htbp]
\centering
  \includegraphics[width=0.48\textwidth]{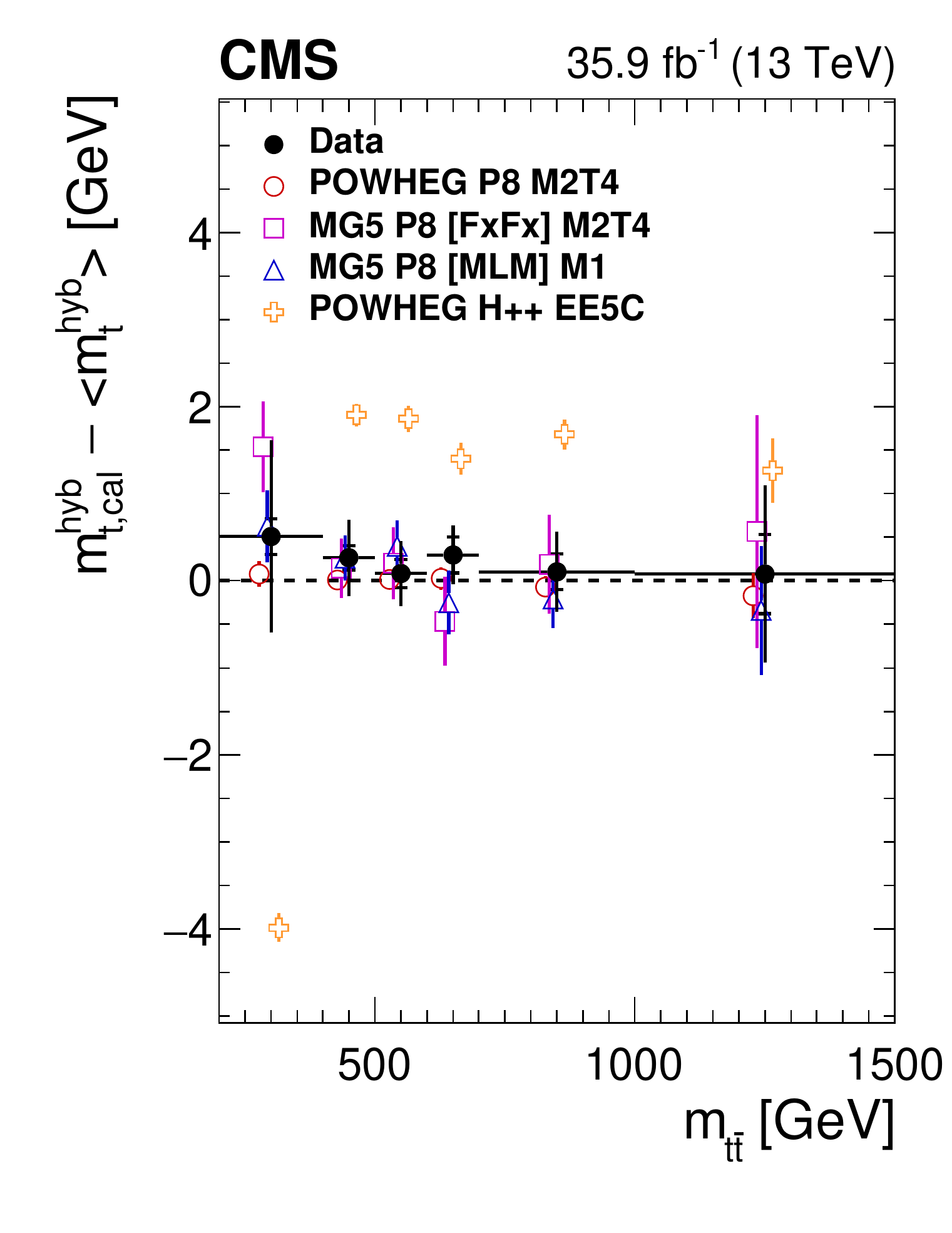}
  \includegraphics[width=0.48\textwidth]{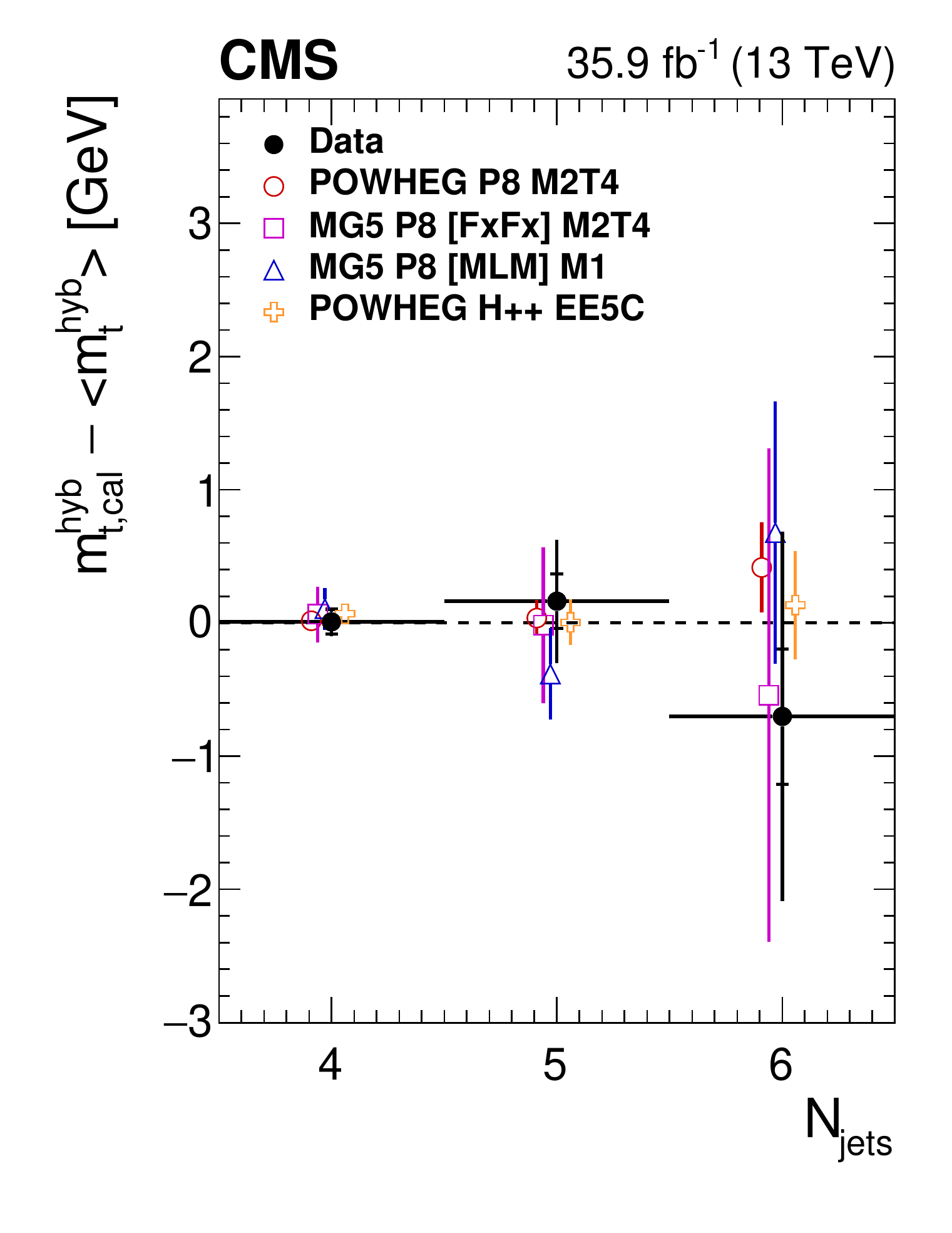}\\
  \includegraphics[width=0.48\textwidth]{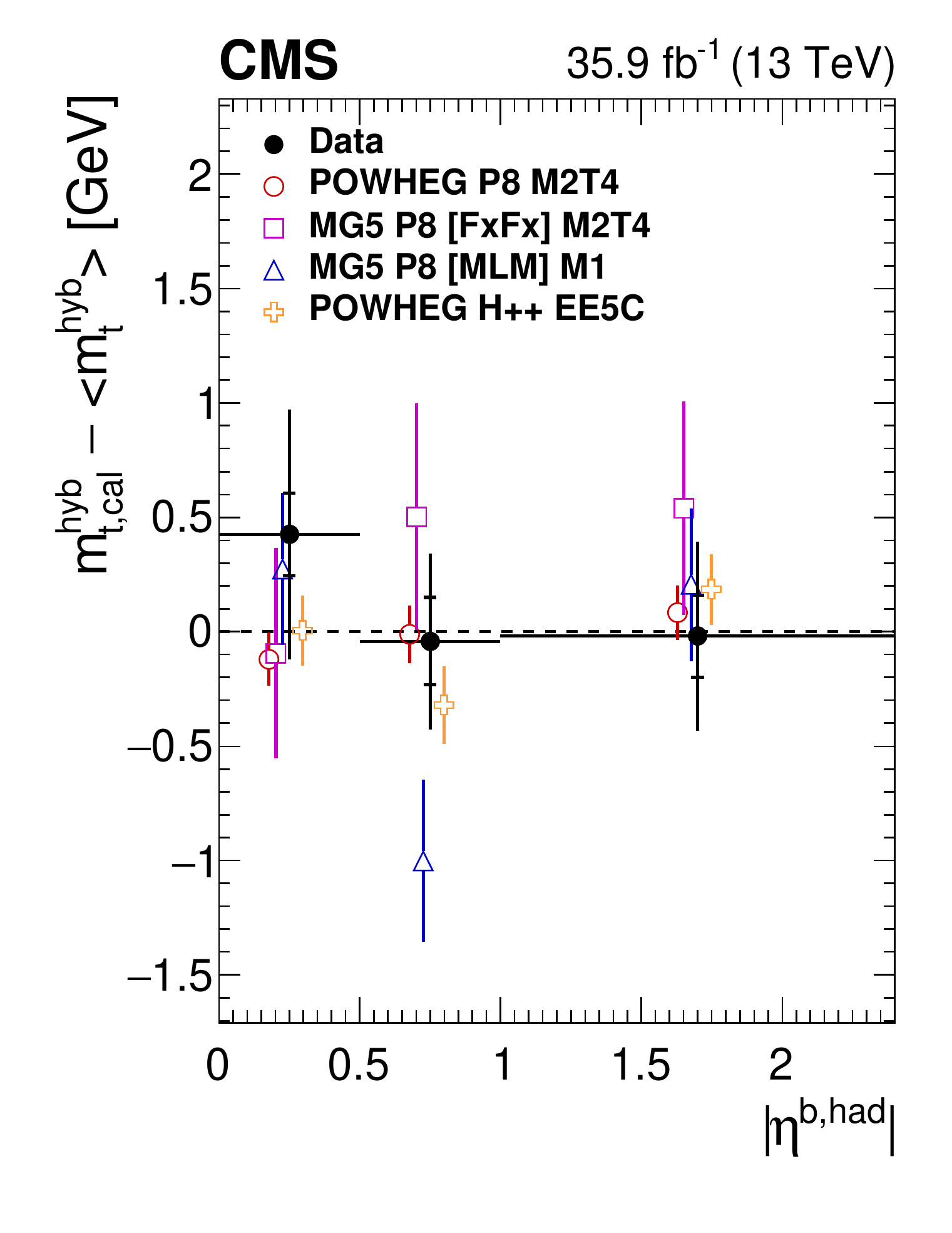}
  \includegraphics[width=0.48\textwidth]{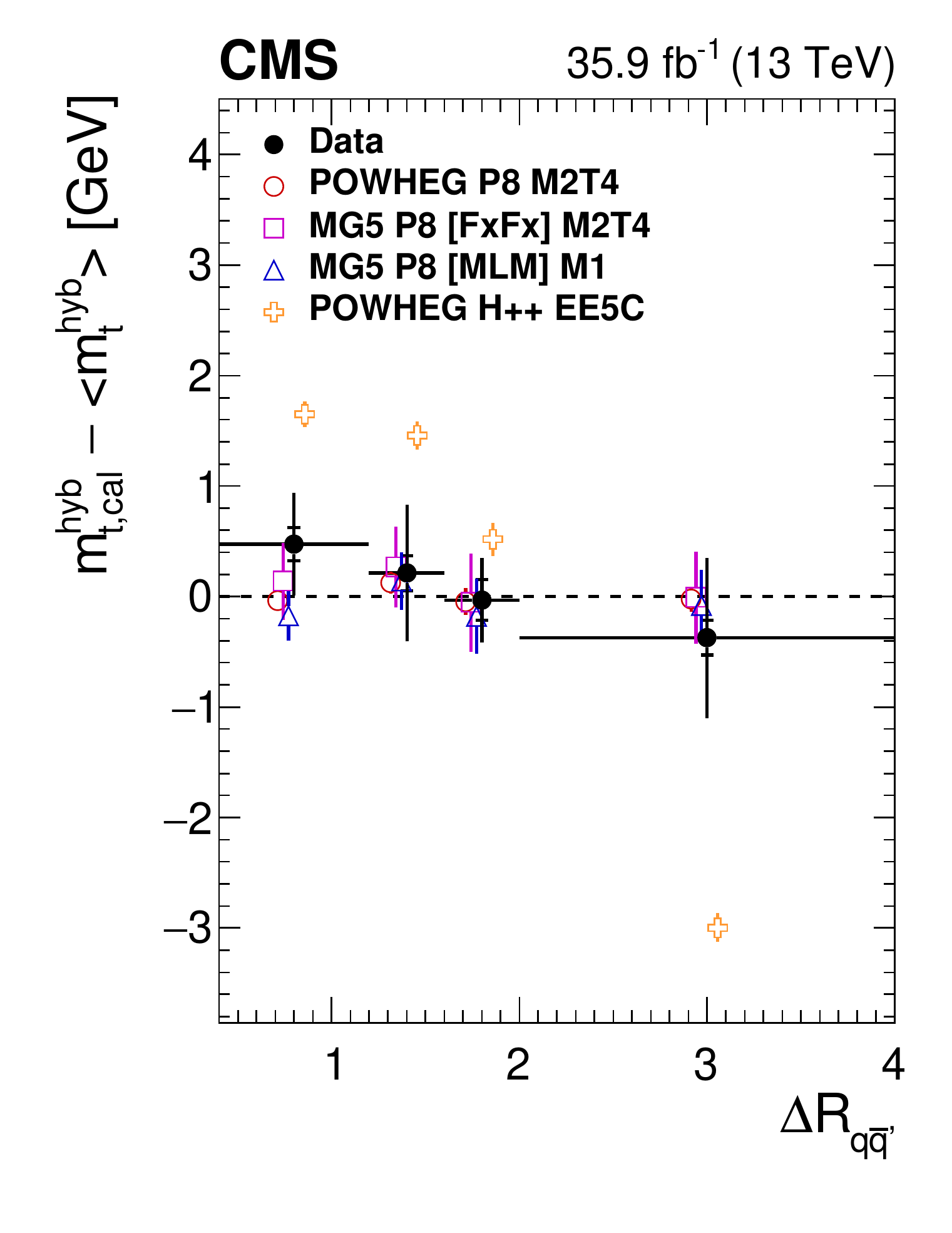}
  \caption{
Measurements of \mtop as a function of the invariant mass of the \ttbar system $m_{\ttbar}$ (upper left), the number of jets $N_\text{jets}$ (upper right), the pseudorapidity of the \PQb{} jet assigned to the hadronic decay branch $\abs{\eta^{\PQb,\text{had}}}$ (lower left) and the $\Delta R$ between the light-quark jets $\Delta R_{\qqbarpr}$ (lower right) compared to different generator models. \diffcaption
}
  \label{fig:diff-gen}
\end{figure*}

\begin{figure*}[!htbp]
\centering
  \includegraphics[width=0.48\textwidth]{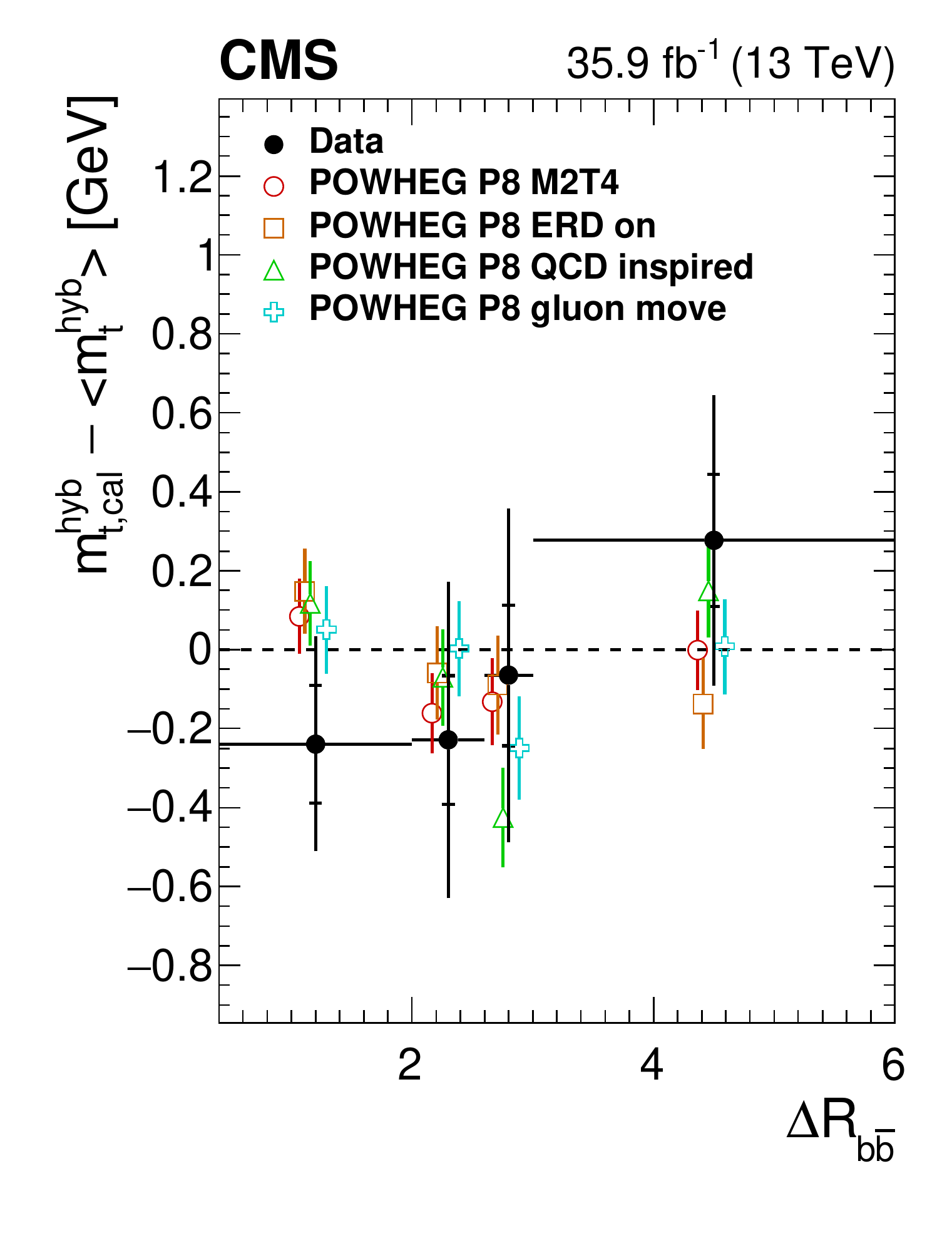}
  \includegraphics[width=0.48\textwidth]{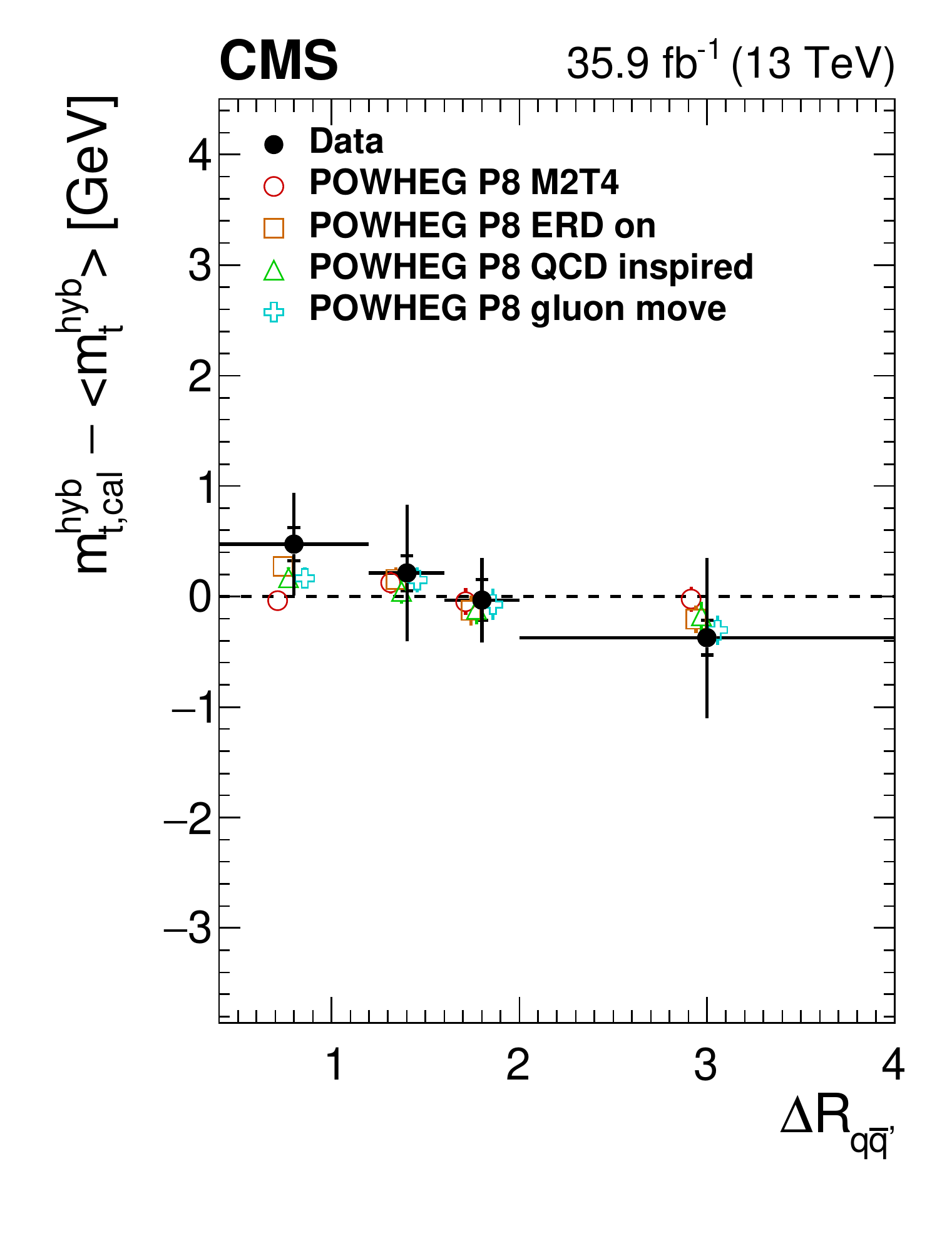}
  \caption{
Measurements of \mtop as a function of the $\Delta R$ between the \PQb{} jets $\Delta R_{\bbbar}$ (left) and the light-quark jets $\Delta R_{\qqbarpr}$ (right) compared to alternative models of color reconnection. The symbols and conventions are the same as in Fig.~\ref{fig:diff-gen}.
}
  \label{fig:diff-cr}
\end{figure*}

No significant deviation of the measured \mtop is observed for the default generator setup of \POWHEG \textsc{p8} M2T4 and there is no evidence for a bias in the measurement.
Only \POWHEG \textsc{h++} EE5C differs from data and all other setups for the dependence of the mass measurement on the invariant mass of the \ttbar system, the \pt of the \PQb{} jet assigned to the hadronic decay branch, and the $\Delta R$ between the light-quark jets.
Figure~\ref{fig:diff-gen} shows the results for $m_{\ttbar}$, $N_\text{jets}$, $\abs{\eta^{ \PQb,\text{had}}}$  and $\Delta R_{\qqbarpr}$ for the different generator setups for the modeling of perturbative QCD.
The large deviations confirm that the \POWHEG~v2 + \HERWIGpp setup without ME corrections to the top quark decay needs improvements to describe the data.
A bias in the measurement of the top quark mass can be spotted by a failure of the model to reproduce differential measurements.
For the color-reconnection models, the $\Delta R_{\bbbar}$ and $\Delta R_{\qqbarpr}$ variables should offer the best sensitivity to the modeling of the color flow.
The comparison is shown in Fig.~\ref{fig:diff-cr}, but the uncertainties in the measurements are too large to rule out any of the different models.


\section{Summary}
\label{Sec:summary}
This study measured the mass of the top quark  using the 2016 data at $\sqrt{s}=13$\TeV  corresponding to an integrated luminosity of 35.9\fbinv, and \POWHEG~v2 interfaced with \PYTHIA~8 with the CUETP8M2T4 tune for the simulation.
The top quark mass is measured to be $172.25\pm 0.08\statJSF\pm 0.62\syst\GeV$ from the selected lepton+jets events.
The result is consistent with the CMS measurements of Run~1  of the LHC at $\sqrt{s}=7$ and 8\TeV, with no shift observed from the new experimental setup and the use of the next-to-leading-order matrix-element generator and the new parton-shower simulation and tune.
Along with the new generator setup, a more advanced treatment of the modeling uncertainties with respect to the Run~1 analysis is employed.
In particular, a broader set of color-reconnection models is considered.
The top quark mass has also been studied as a function of the event-level kinematic properties, and no indications of a bias in the measurements are observed.

\begin{acknowledgments}
We congratulate our colleagues in the CERN accelerator departments for the excellent performance of the LHC and thank the technical and administrative staffs at CERN and at other CMS institutes for their contributions to the success of the CMS effort. In addition, we gratefully acknowledge the computing centers and personnel of the Worldwide LHC Computing Grid for delivering so effectively the computing infrastructure essential to our analyses. Finally, we acknowledge the enduring support for the construction and operation of the LHC and the CMS detector provided by the following funding agencies: BMWFW and FWF (Austria); FNRS and FWO (Belgium); CNPq, CAPES, FAPERJ, and FAPESP (Brazil); MES (Bulgaria); CERN; CAS, MoST, and NSFC (China); COLCIENCIAS (Colombia); MSES and CSF (Croatia); RPF (Cyprus); SENESCYT (Ecuador); MoER, ERC IUT, and ERDF (Estonia); Academy of Finland, MEC, and HIP (Finland); CEA and CNRS/IN2P3 (France); BMBF, DFG, and HGF (Germany); GSRT (Greece); NKFIA (Hungary); DAE and DST (India); IPM (Iran); SFI (Ireland); INFN (Italy); MSIP and NRF (Republic of Korea); LAS (Lithuania); MOE and UM (Malaysia); BUAP, CINVESTAV, CONACYT, LNS, SEP, and UASLP-FAI (Mexico); MBIE (New Zealand); PAEC (Pakistan); MSHE and NSC (Poland); FCT (Portugal); JINR (Dubna); MON, RosAtom, RAS and RFBR (Russia); MESTD (Serbia); SEIDI, CPAN, PCTI and FEDER (Spain); Swiss Funding Agencies (Switzerland); MST (Taipei); ThEPCenter, IPST, STAR, and NSTDA (Thailand); TUBITAK and TAEK (Turkey); NASU and SFFR (Ukraine); STFC (United Kingdom); DOE and NSF (USA).

\hyphenation{Rachada-pisek} Individuals have received support from the Marie-Curie program and the European Research Council and Horizon 2020 Grant, contract No. 675440 (European Union); the Leventis Foundation; the A. P. Sloan Foundation; the Alexander von Humboldt Foundation; the Belgian Federal Science Policy Office; the Fonds pour la Formation \`a la Recherche dans l'Industrie et dans l'Agriculture (FRIA-Belgium); the Agentschap voor Innovatie door Wetenschap en Technologie (IWT-Belgium); the F.R.S.-FNRS and FWO (Belgium) under the ``Excellence of Science - EOS" - be.h project n. 30820817; the Ministry of Education, Youth and Sports (MEYS) of the Czech Republic; the Lend\"ulet (``Momentum") Programme and the J\'anos Bolyai Research Scholarship of the Hungarian Academy of Sciences, the New National Excellence Program \'UNKP, the NKFIA research grants 123842, 123959, 124845, 124850 and 125105 (Hungary); the Council of Science and Industrial Research, India; the HOMING PLUS program of the Foundation for Polish Science, cofinanced from European Union, Regional Development Fund, the Mobility Plus program of the Ministry of Science and Higher Education, the National Science Center (Poland), contracts Harmonia 2014/14/M/ST2/00428, Opus 2014/13/B/ST2/02543, 2014/15/B/ST2/03998, and 2015/19/B/ST2/02861, Sonata-bis 2012/07/E/ST2/01406; the National Priorities Research Program by Qatar National Research Fund; the Programa Estatal de Fomento de la Investigaci{\'o}n Cient{\'i}fica y T{\'e}cnica de Excelencia Mar\'{\i}a de Maeztu, grant MDM-2015-0509 and the Programa Severo Ochoa del Principado de Asturias; the Thalis and Aristeia programs cofinanced by EU-ESF and the Greek NSRF; the Rachadapisek Sompot Fund for Postdoctoral Fellowship, Chulalongkorn University and the Chulalongkorn Academic into Its 2nd Century Project Advancement Project (Thailand); the Welch Foundation, contract C-1845; and the Weston Havens Foundation (USA).
\end{acknowledgments}

\bibliography{auto_generated}

\providecommand{\href}[2]{#2}\begingroup\raggedright\begin{thebibliography}{10}%
\makeatletter
\providecommand{\hrefCMSnoop }[0]{\@secondoftwo}%
\makeatother
\providecommand{\doi}{\texttt{doi:}\begingroup \urlstyle{tt}\Url}

\bibitem{Baak:2014ora}
M.~Baak\hrefCMSnoop {}{ {et~al.}, ``{The global electroweak fit at NNLO and
  prospects for the LHC and ILC}'',} \textit{ Eur. Phys. J. C} \textbf{ 74}
  (2014) 3046,
  \href{http://dx.doi.org/10.1140/epjc/s10052-014-3046-5}{\doi{10.1140/epjc/s10052-014-3046-5}},
\href{http://www.arXiv.org/abs/1407.3792}{\texttt{arXiv:1407.3792}}.

\bibitem{Degrassi:2012ry}
G.~Degrassi\hrefCMSnoop {}{ {et~al.}, ``{Higgs mass and vacuum stability in the
  Standard Model at NNLO}'',} \textit{ JHEP} \textbf{ 08} (2012) 098,
  \href{http://dx.doi.org/10.1007/JHEP08(2012)098}{\doi{10.1007/JHEP08(2012)098}},
\href{http://www.arXiv.org/abs/1205.6497}{\texttt{arXiv:1205.6497}}.

\bibitem{Olive:2016xmw}
\hrefCMSnoop {}{{Particle Data Group}, C.~Patrignani {et~al.}, ``Review of
  particle physics'',} \textit{ Chin. Phys. C} \textbf{ 40} (2016) 100001,
  \href{http://dx.doi.org/10.1088/1674-1137/40/10/100001}{\doi{10.1088/1674-1137/40/10/100001}}.

\bibitem{TevatronElectroweakWorkingGroup:2016lid}
\hrefCMSnoop {}{{CDF and \DZERO Collaborations}, ``{Combination of CDF and D0
  results on the mass of the top quark using up to $9.7\:{\rm fb}^{-1}$ at the
  Tevatron}'',} (2016).
\href{http://www.arXiv.org/abs/1608.01881}{\texttt{arXiv:1608.01881}}.

\bibitem{Aaboud:2016igd}
\hrefCMSnoop {}{{ATLAS Collaboration}, ``{Measurement of the top quark mass in
  the $\ttbar\to$ dilepton channel from $\sqrt{s}=8$ TeV ATLAS data}'',}
  \textit{ Phys. Lett. B} \textbf{ 761} (2016) 350,
  \href{http://dx.doi.org/10.1016/j.physletb.2016.08.042}{\doi{10.1016/j.physletb.2016.08.042}},
\href{http://www.arXiv.org/abs/1606.02179}{\texttt{arXiv:1606.02179}}.

\bibitem{Khachatryan:2015hba}
\hrefCMSnoop {}{{CMS Collaboration}, ``{Measurement of the top quark mass using
  proton-proton data at ${\sqrt{s}} = 7$ and 8 TeV}'',} \textit{ Phys. Rev. D}
  \textbf{ 93} (2016) 072004,
  \href{http://dx.doi.org/10.1103/PhysRevD.93.072004}{\doi{10.1103/PhysRevD.93.072004}},
\href{http://www.arXiv.org/abs/1509.04044}{\texttt{arXiv:1509.04044}}.

\bibitem{Hoang:2008xm}
\hrefCMSnoop {}{A.~H. Hoang and I.~W. Stewart, ``Top mass measurements from
  jets and the {Tevatron} top-quark mass'',} \textit{ Nucl. Phys. Proc. Suppl.}
  \textbf{ 185} (2008) 220,
  \href{http://dx.doi.org/10.1016/j.nuclphysbps.2008.10.028}{\doi{10.1016/j.nuclphysbps.2008.10.028}},
\href{http://www.arXiv.org/abs/0808.0222}{\texttt{arXiv:0808.0222}}.

\bibitem{Moch:2014tta}
\hrefCMSnoop {}{S.~Moch {et~al.}, ``{High precision fundamental constants at
  the TeV scale}'',} (2014).
\href{http://www.arXiv.org/abs/1405.4781}{\texttt{arXiv:1405.4781}}.

\bibitem{Marquard:2015qpa}
\hrefCMSnoop {}{P.~Marquard, A.~V. Smirnov, V.~A. Smirnov, and M.~Steinhauser,
  ``Quark mass relations to four-loop order in perturbative {QCD}'',} \textit{
  Phys. Rev. Lett.} \textbf{ 114} (2015) 142002,
  \href{http://dx.doi.org/10.1103/PhysRevLett.114.142002}{\doi{10.1103/PhysRevLett.114.142002}},
\href{http://www.arXiv.org/abs/1502.01030}{\texttt{arXiv:1502.01030}}.

\bibitem{Beneke:2016cbu}
\hrefCMSnoop {}{M.~Beneke, P.~Marquard, P.~Nason, and M.~Steinhauser, ``{On the
  ultimate uncertainty of the top quark pole mass}'',} \textit{ Phys. Lett. B}
  \textbf{ 775} (2017) 63,
  \href{http://dx.doi.org/10.1016/j.physletb.2017.10.054}{\doi{10.1016/j.physletb.2017.10.054}},
\href{http://www.arXiv.org/abs/1605.03609}{\texttt{arXiv:1605.03609}}.

\bibitem{Butenschoen:2016lpz}
M.~Butenschoen\hrefCMSnoop {}{ {et~al.}, ``Top quark mass calibration for
  {Monte Carlo} event generators'',} \textit{ Phys. Rev. Lett.} \textbf{ 117}
  (2016) 232001,
  \href{http://dx.doi.org/10.1103/PhysRevLett.117.232001}{\doi{10.1103/PhysRevLett.117.232001}},
\href{http://www.arXiv.org/abs/1608.01318}{\texttt{arXiv:1608.01318}}.

\bibitem{Hoang:2017btd}
\hrefCMSnoop {}{A.~H. Hoang, C.~Lepenik, and M.~Preisser, ``On the light
  massive flavor dependence of the large order asymptotic behavior and the
  ambiguity of the pole mass'',} \textit{ JHEP} \textbf{ 09} (2017) 099,
  \href{http://dx.doi.org/10.1007/JHEP09(2017)099}{\doi{10.1007/JHEP09(2017)099}},
\href{http://www.arXiv.org/abs/1706.08526}{\texttt{arXiv:1706.08526}}.

\bibitem{Nason:2018qkf}
\href
  {http://inspirehep.net/record/1648157/files/arXiv:1801.04826.pdf}{P.~Nason,
  ``The top quark mass at the {LHC}'',} in \textit{ {10th International
  Workshop on Top Quark Physics (TOP2017) Braga, Portugal, September 17-22,
  2017}}.
\newblock 2018.
\newblock
\href{http://www.arXiv.org/abs/1801.04826}{\texttt{arXiv:1801.04826}}.
\newblock

\bibitem{Chatrchyan:2008zzk}
\hrefCMSnoop {}{{{CMS}} Collaboration, ``The {CMS} experiment at the {CERN
  LHC}'',} \textit{ JINST} \textbf{ 3} (2008) S08004,
\href{http://dx.doi.org/10.1088/1748-0221/3/08/S08004}{\doi{10.1088/1748-0221/3/08/S08004}}.

\bibitem{Sirunyan:2017ulk}
\hrefCMSnoop {}{{CMS Collaboration}, ``{Particle-flow reconstruction and global
  event description with the CMS detector}'',} \textit{ JINST} \textbf{ 12}
  (2017) P10003,
  \href{http://dx.doi.org/10.1088/1748-0221/12/10/P10003}{\doi{10.1088/1748-0221/12/10/P10003}},
\href{http://www.arXiv.org/abs/1706.04965}{\texttt{arXiv:1706.04965}}.

\bibitem{Cacciari:2005hq}
\hrefCMSnoop {}{M.~Cacciari and G.~P. Salam, ``{Dispelling the $N^{3}$ myth for
  the $k_{\rm t}$ jet-finder}'',} \textit{ Phys. Lett. B} \textbf{ 641} (2006)
  57,
  \href{http://dx.doi.org/10.1016/j.physletb.2006.08.037}{\doi{10.1016/j.physletb.2006.08.037}},
\href{http://www.arXiv.org/abs/hep-ph/0512210}{\texttt{arXiv:hep-ph/0512210}}.

\bibitem{Cacciari:2008gp}
\hrefCMSnoop {}{M.~Cacciari, G.~P. Salam, and G.~Soyez, ``{The anti-$k_{\rm t}$
  jet clustering algorithm}'',} \textit{ JHEP} \textbf{ 04} (2008) 063,
  \href{http://dx.doi.org/10.1088/1126-6708/2008/04/063}{\doi{10.1088/1126-6708/2008/04/063}},
\href{http://www.arXiv.org/abs/0802.1189}{\texttt{arXiv:0802.1189}}.

\bibitem{Cacciari:2011ma}
\hrefCMSnoop {}{M.~Cacciari, G.~P. Salam, and G.~Soyez, ``{FastJet user
  manual}'',} \textit{ Eur. Phys. J. C} \textbf{ 72} (2012) 1896,
  \href{http://dx.doi.org/10.1140/epjc/s10052-012-1896-2}{\doi{10.1140/epjc/s10052-012-1896-2}},
\href{http://www.arXiv.org/abs/1111.6097}{\texttt{arXiv:1111.6097}}.

\bibitem{Cacciari:2007fd}
\hrefCMSnoop {}{M.~Cacciari and G.~P. Salam, ``{Pileup subtraction using jet
  areas}'',} \textit{ Phys. Lett. B} \textbf{ 659} (2008) 119,
  \href{http://dx.doi.org/10.1016/j.physletb.2007.09.077}{\doi{10.1016/j.physletb.2007.09.077}},
\href{http://www.arXiv.org/abs/0707.1378}{\texttt{arXiv:0707.1378}}.

\bibitem{Khachatryan:2016kdb}
\hrefCMSnoop {}{{CMS Collaboration}, ``Jet energy scale and resolution in the
  {CMS} experiment in pp collisions at 8\,{TeV}'',} \textit{ JINST} \textbf{
  12} (2017) P02014,
  \href{http://dx.doi.org/10.1088/1748-0221/12/02/P02014}{\doi{10.1088/1748-0221/12/02/P02014}},
\href{http://www.arXiv.org/abs/1607.03663}{\texttt{arXiv:1607.03663}}.

\bibitem{CMS-PAS-JME-16-003}
\href {http://cds.cern.ch/record/2256875}{{CMS Collaboration}, ``{Jet
  algorithms performance in 13 TeV data}'',} CMS Physics Analysis Summary
  CMS-PAS-JME-16-003, 2017.

\bibitem{CMS-PAS-LUM-17-001}
\href {http://cds.cern.ch/record/2257069}{{CMS Collaboration}, ``{CMS}
  luminosity measurements for the 2016 data taking period'',} CMS Physics
  Analysis Summary CMS-PAS-LUM-17-001, 2017.

\bibitem{Nason:2004rx}
\hrefCMSnoop {}{P.~Nason, ``{A new method for combining NLO QCD with shower
  Monte Carlo algorithms}'',} \textit{ JHEP} \textbf{ 11} (2004) 040,
  \href{http://dx.doi.org/10.1088/1126-6708/2004/11/040}{\doi{10.1088/1126-6708/2004/11/040}},
  \href{http://www.arXiv.org/abs/hep-ph/0409146}{\texttt{arXiv:hep-ph/0409146}}.

\bibitem{Frixione:2007vw}
\hrefCMSnoop {}{S.~Frixione, P.~Nason, and C.~Oleari, ``{Matching NLO QCD
  computations with parton shower simulations: the \POWHEG method}'',} \textit{
  JHEP} \textbf{ 11} (2007) 070,
  \href{http://dx.doi.org/10.1088/1126-6708/2007/11/070}{\doi{10.1088/1126-6708/2007/11/070}},
  \href{http://www.arXiv.org/abs/0709.2092}{\texttt{arXiv:0709.2092}}.

\bibitem{Alioli:2010xd}
\hrefCMSnoop {}{S.~Alioli, P.~Nason, C.~Oleari, and E.~Re, ``{A general
  framework for implementing NLO calculations in shower Monte Carlo programs:
  the \POWHEG BOX}'',} \textit{ JHEP} \textbf{ 06} (2010) 043,
  \href{http://dx.doi.org/10.1007/JHEP06(2010)043}{\doi{10.1007/JHEP06(2010)043}},
  \href{http://www.arXiv.org/abs/1002.2581}{\texttt{arXiv:1002.2581}}.

\bibitem{Campbell:2014kua}
\hrefCMSnoop {}{J.~M. Campbell, R.~K. Ellis, P.~Nason, and E.~Re, ``{Top-pair
  production and decay at NLO matched with parton showers}'',} \textit{ JHEP}
  \textbf{ 04} (2015) 114,
  \href{http://dx.doi.org/10.1007/JHEP04(2015)114}{\doi{10.1007/JHEP04(2015)114}},
\href{http://www.arXiv.org/abs/1412.1828}{\texttt{arXiv:1412.1828}}.

\bibitem{Sjostrand:2007gs}
\hrefCMSnoop {}{T.~Sj{\"o}strand, S.~Mrenna, and P.~Skands, ``{A brief
  introduction to \PYTHIA 8.1}'',} \textit{ Comput. Phys. Commun.} \textbf{
  178} (2008) 852,
  \href{http://dx.doi.org/10.1016/j.cpc.2008.01.036}{\doi{10.1016/j.cpc.2008.01.036}},
\href{http://www.arXiv.org/abs/0710.3820}{\texttt{arXiv:0710.3820}}.

\bibitem{Skands:2014pea}
\hrefCMSnoop {}{P.~Skands, S.~Carrazza, and J.~Rojo, ``{Tuning \PYTHIA 8.1: the
  Monash 2013 tune}'',} \textit{ Eur. Phys. J. C} \textbf{ 74} (2014) 3024,
  \href{http://dx.doi.org/10.1140/epjc/s10052-014-3024-y}{\doi{10.1140/epjc/s10052-014-3024-y}},
\href{http://www.arXiv.org/abs/1404.5630}{\texttt{arXiv:1404.5630}}.

\bibitem{CMS-PAS-TOP-16-021}
\href {https://cds.cern.ch/record/2235192}{{CMS Collaboration},
  ``{Investigations of the impact of the parton shower tuning in \PYTHIA~8 in
  the modelling of $\mathrm{t\overline{t}}$ at $\sqrt{s}=8$ and 13 TeV}'',} CMS
  Physics Analysis Summary CMS-PAS-TOP-16-021, 2016.

\bibitem{Alioli:2009je}
\hrefCMSnoop {}{S.~Alioli, P.~Nason, C.~Oleari, and E.~Re, ``{NLO single-top
  production matched with shower in \POWHEG: $s$- and $t$-channel
  contributions}'',} \textit{ JHEP} \textbf{ 09} (2009) 111,
  \href{http://dx.doi.org/10.1088/1126-6708/2009/09/111}{\doi{10.1088/1126-6708/2009/09/111}},
  \href{http://www.arXiv.org/abs/0907.4076}{\texttt{arXiv:0907.4076}}.
  [Erratum: \DOI{10.1007/JHEP02(2010)011}].

\bibitem{Re:2010bp}
\hrefCMSnoop {}{E.~Re, ``{Single-top Wt-channel production matched with parton
  showers using the \POWHEG method}'',} \textit{ Eur. Phys. J. C} \textbf{ 71}
  (2011) 1547,
  \href{http://dx.doi.org/10.1140/epjc/s10052-011-1547-z}{\doi{10.1140/epjc/s10052-011-1547-z}},
  \href{http://www.arXiv.org/abs/1009.2450}{\texttt{arXiv:1009.2450}}.

\bibitem{Alwall:2014hca}
J.~Alwall\hrefCMSnoop {}{ {et~al.}, ``{The automated computation of tree-level
  and next-to-leading order differential cross sections, and their matching to
  parton shower simulations}'',} \textit{ JHEP} \textbf{ 07} (2014) 079,
  \href{http://dx.doi.org/10.1007/JHEP07(2014)079}{\doi{10.1007/JHEP07(2014)079}},
\href{http://www.arXiv.org/abs/1405.0301}{\texttt{arXiv:1405.0301}}.

\bibitem{Alwall:2007fs}
J.~Alwall\hrefCMSnoop {}{ {et~al.}, ``{Comparative study of various algorithms
  for the merging of parton showers and matrix elements in hadronic
  collisions}'',} \textit{ Eur. Phys. J. C} \textbf{ 53} (2008) 473,
  \href{http://dx.doi.org/10.1140/epjc/s10052-007-0490-5}{\doi{10.1140/epjc/s10052-007-0490-5}},
\href{http://www.arXiv.org/abs/0706.2569}{\texttt{arXiv:0706.2569}}.

\bibitem{Frederix:2012ps}
\hrefCMSnoop {}{R.~Frederix and S.~Frixione, ``{Merging meets matching in
  MC@NLO}'',} \textit{ JHEP} \textbf{ 12} (2012) 061,
  \href{http://dx.doi.org/10.1007/JHEP12(2012)061}{\doi{10.1007/JHEP12(2012)061}},
\href{http://www.arXiv.org/abs/1209.6215}{\texttt{arXiv:1209.6215}}.

\bibitem{Ball:2014uwa}
\hrefCMSnoop {}{{NNPDF} Collaboration, ``{Parton distributions for the LHC Run
  II}'',} \textit{ JHEP} \textbf{ 04} (2015) 040,
  \href{http://dx.doi.org/10.1007/JHEP04(2015)040}{\doi{10.1007/JHEP04(2015)040}},
\href{http://www.arXiv.org/abs/1410.8849}{\texttt{arXiv:1410.8849}}.

\bibitem{Czakon:2011xx}
\hrefCMSnoop {}{M.~Czakon and A.~Mitov, ``{Top++: A program for the calculation
  of the top-pair cross-section at hadron colliders}'',} \textit{ Comput. Phys.
  Commun.} \textbf{ 185} (2014) 2930,
  \href{http://dx.doi.org/10.1016/j.cpc.2014.06.021}{\doi{10.1016/j.cpc.2014.06.021}},
\href{http://www.arXiv.org/abs/1112.5675}{\texttt{arXiv:1112.5675}}.

\bibitem{Li:2012wna}
\hrefCMSnoop {}{Y.~Li and F.~Petriello, ``{Combining QCD and electroweak
  corrections to dilepton production in FEWZ}'',} \textit{ Phys. Rev. D}
  \textbf{ 86} (2012) 094034,
  \href{http://dx.doi.org/10.1103/PhysRevD.86.094034}{\doi{10.1103/PhysRevD.86.094034}},
\href{http://www.arXiv.org/abs/1208.5967}{\texttt{arXiv:1208.5967}}.

\bibitem{Kant:2014oha}
P.~Kant\hrefCMSnoop {}{ {et~al.}, ``{HatHor for single top-quark production:
  Updated predictions and uncertainty estimates for single top-quark production
  in hadronic collisions}'',} \textit{ Comput. Phys. Commun.} \textbf{ 191}
  (2015) 74,
  \href{http://dx.doi.org/10.1016/j.cpc.2015.02.001}{\doi{10.1016/j.cpc.2015.02.001}},
\href{http://www.arXiv.org/abs/1406.4403}{\texttt{arXiv:1406.4403}}.

\bibitem{Kidonakis:2012rm}
\hrefCMSnoop {}{N.~Kidonakis, ``{NNLL threshold resummation for top-pair and
  single-top production}'',} \textit{ Phys. Part. Nucl.} \textbf{ 45} (2014)
  714,
  \href{http://dx.doi.org/10.1134/S1063779614040091}{\doi{10.1134/S1063779614040091}},
\href{http://www.arXiv.org/abs/1210.7813}{\texttt{arXiv:1210.7813}}.

\bibitem{Agostinelli:2002hh}
\hrefCMSnoop {}{{GEANT4} Collaboration, ``{GEANT4}---a simulation toolkit'',}
  \textit{ Nucl. Instrum. Meth. A} \textbf{ 506} (2003) 250,
\href{http://dx.doi.org/10.1016/S0168-9002(03)01368-8}{\doi{10.1016/S0168-9002(03)01368-8}}.

\bibitem{Chatrchyan:2012xi}
\hrefCMSnoop {}{{CMS Collaboration}, ``{Performance of CMS muon reconstruction
  in pp collision events at $\sqrt{s} = 7$\TeV}'',} \textit{ JINST} \textbf{ 7}
  (2012) P10002,
  \href{http://dx.doi.org/10.1088/1748-0221/7/10/P10002}{\doi{10.1088/1748-0221/7/10/P10002}},
\href{http://www.arXiv.org/abs/1206.4071}{\texttt{arXiv:1206.4071}}.

\bibitem{Khachatryan:2015hwa}
\hrefCMSnoop {}{{CMS Collaboration}, ``{Performance of electron reconstruction
  and selection with the CMS detector in proton-proton collisions at
  $\sqrt{s}=8~\mathrm{TeV}$}'',} \textit{ JINST} \textbf{ 10} (2015) P06005,
  \href{http://dx.doi.org/10.1088/1748-0221/10/06/P06005}{\doi{10.1088/1748-0221/10/06/P06005}},
\href{http://www.arXiv.org/abs/1502.02701}{\texttt{arXiv:1502.02701}}.

\bibitem{Sirunyan:2017ezt}
\hrefCMSnoop {}{{CMS Collaboration}, ``Identification of heavy-flavour jets
  with the {CMS} detector in pp collisions at 13 {TeV}'',} \textit{ JINST}
  \textbf{ 13} (2018) P05011,
  \href{http://dx.doi.org/10.1088/1748-0221/13/05/P05011}{\doi{10.1088/1748-0221/13/05/P05011}},
\href{http://www.arXiv.org/abs/1712.07158}{\texttt{arXiv:1712.07158}}.

\bibitem{Abbott:1998dc}
\hrefCMSnoop {}{{\DZERO} Collaboration, ``{Direct measurement of the top quark
  mass at \DZERO}'',} \textit{ Phys. Rev. D} \textbf{ 58} (1998) 052001,
  \href{http://dx.doi.org/10.1103/PhysRevD.58.052001}{\doi{10.1103/PhysRevD.58.052001}},
\href{http://www.arXiv.org/abs/hep-ex/9801025}{\texttt{arXiv:hep-ex/9801025}}.

\bibitem{Abdallah:2008xh}
\hrefCMSnoop {}{{DELPHI} Collaboration, ``{Measurement of the mass and width of
  the W boson in ${\rm e}^{+}{\rm e}^{-}$ collisions at $\sqrt{s}$ = 161 -- 209
  GeV}'',} \textit{ Eur. Phys. J. C} \textbf{ 55} (2008) 1,
  \href{http://dx.doi.org/10.1140/epjc/s10052-008-0585-7}{\doi{10.1140/epjc/s10052-008-0585-7}},
\href{http://www.arXiv.org/abs/0803.2534}{\texttt{arXiv:0803.2534}}.

\bibitem{Chatrchyan:2012cz}
\hrefCMSnoop {}{{CMS Collaboration}, ``{Measurement of the top-quark mass in
  $\ttbar$ events with lepton+jets final states in pp collisions at
  $\sqrt{s}=7$ TeV}'',} \textit{ JHEP} \textbf{ 12} (2012) 105,
  \href{http://dx.doi.org/10.1007/JHEP12(2012)105}{\doi{10.1007/JHEP12(2012)105}},
\href{http://www.arXiv.org/abs/1209.2319}{\texttt{arXiv:1209.2319}}.

\bibitem{Argyropoulos:2014zoa}
\hrefCMSnoop {}{S.~Argyropoulos and T.~Sj{\"o}strand, ``{Effects of color
  reconnection on $\ttbar$ final states at the LHC}'',} \textit{ JHEP} \textbf{
  11} (2014) 043,
  \href{http://dx.doi.org/10.1007/JHEP11(2014)043}{\doi{10.1007/JHEP11(2014)043}},
\href{http://www.arXiv.org/abs/1407.6653}{\texttt{arXiv:1407.6653}}.

\bibitem{Christiansen:2015yqa}
\hrefCMSnoop {}{J.~R. Christiansen and P.~Skands, ``{String formation beyond
  leading colour}'',} \textit{ JHEP} \textbf{ 08} (2015) 003,
  \href{http://dx.doi.org/10.1007/JHEP08(2015)003}{\doi{10.1007/JHEP08(2015)003}},
\href{http://www.arXiv.org/abs/1505.01681}{\texttt{arXiv:1505.01681}}.

\bibitem{CMS-PAS-JME-15-001}
\href {http://cds.cern.ch/record/2104039}{{ATLAS and CMS Collaborations},
  ``{Jet energy scale uncertainty correlations between ATLAS and CMS at 8
  TeV}'',} {} ATL-PHYS-PUB-2015-049, CMS-PAS-JME-15-001, 2015.

\bibitem{Khachatryan:2014iya}
\hrefCMSnoop {}{{CMS Collaboration}, ``{Measurement of the $t$-channel
  single-top-quark production cross section and of the $\mid V_{tb} \mid$ CKM
  matrix element in pp collisions at $\sqrt{s}$= 8 TeV}'',} \textit{ JHEP}
  \textbf{ 06} (2014) 090,
  \href{http://dx.doi.org/10.1007/JHEP06(2014)090}{\doi{10.1007/JHEP06(2014)090}},
\href{http://www.arXiv.org/abs/1403.7366}{\texttt{arXiv:1403.7366}}.

\bibitem{Sirunyan:2016cdg}
\hrefCMSnoop {}{{CMS Collaboration}, ``{Cross section measurement of
  $t$-channel single top quark production in pp collisions at $\sqrt s =$ 13
  TeV}'',} \textit{ Phys. Lett. B} \textbf{ 772} (2017) 752,
  \href{http://dx.doi.org/10.1016/j.physletb.2017.07.047}{\doi{10.1016/j.physletb.2017.07.047}},
\href{http://www.arXiv.org/abs/1610.00678}{\texttt{arXiv:1610.00678}}.

\bibitem{Khachatryan:2016ipq}
\hrefCMSnoop {}{{CMS Collaboration}, ``{Measurement of the production cross
  section of a W boson in association with two b jets in pp collisions at
  $\sqrt{s} = 8{\,\mathrm{{TeV}}} $}'',} \textit{ Eur. Phys. J C.} \textbf{ 77}
  (2017) 92,
  \href{http://dx.doi.org/10.1140/epjc/s10052-016-4573-z}{\doi{10.1140/epjc/s10052-016-4573-z}},
\href{http://www.arXiv.org/abs/1608.07561}{\texttt{arXiv:1608.07561}}.

\bibitem{Khachatryan:2016iob}
\hrefCMSnoop {}{{CMS Collaboration}, ``{Measurements of the associated
  production of a Z boson and b jets in pp collisions at ${\sqrt{s}} = 8\,\text
  {TeV} $}'',} \textit{ Eur. Phys. J. C} \textbf{ 77} (2017) 751,
  \href{http://dx.doi.org/10.1140/epjc/s10052-017-5140-y}{\doi{10.1140/epjc/s10052-017-5140-y}},
\href{http://www.arXiv.org/abs/1611.06507}{\texttt{arXiv:1611.06507}}.

\bibitem{Khachatryan:2016tgp}
\hrefCMSnoop {}{{CMS Collaboration}, ``Measurement of the {WZ} production cross
  section in pp collisions at $\sqrt(s) =$ 13 {TeV}'',} \textit{ Phys. Lett. B}
  \textbf{ 766} (2017) 268,
  \href{http://dx.doi.org/10.1016/j.physletb.2017.01.011}{\doi{10.1016/j.physletb.2017.01.011}},
\href{http://www.arXiv.org/abs/1607.06943}{\texttt{arXiv:1607.06943}}.

\bibitem{Sirunyan:2017zjc}
\hrefCMSnoop {}{{CMS Collaboration}, ``Measurements of the $\mathrm {p}\mathrm
  {p}\rightarrow \mathrm{Z}\mathrm{Z}$ production cross section and the
  $\mathrm{Z}\rightarrow 4\ell $ branching fraction, and constraints on
  anomalous triple gauge couplings at $\sqrt{s} = 13\,\text {TeV} $'',}
  \textit{ Eur. Phys. J. C} \textbf{ 78} (2018) 165,
  \href{http://dx.doi.org/10.1140/epjc/s10052-018-5567-9}{\doi{10.1140/epjc/s10052-018-5567-9}},
\href{http://www.arXiv.org/abs/1709.08601}{\texttt{arXiv:1709.08601}}.

\bibitem{Sjostrand:2006za}
\hrefCMSnoop {}{T.~{Sj\"ostrand}, S.~Mrenna, and P.~Skands, ``{\PYTHIA 6.4
  physics and manual}'',} \textit{ JHEP} \textbf{ 05} (2006) 026,
  \href{http://dx.doi.org/10.1088/1126-6708/2006/05/026}{\doi{10.1088/1126-6708/2006/05/026}},
\href{http://www.arXiv.org/abs/hep-ph/0603175}{\texttt{arXiv:hep-ph/0603175}}.

\bibitem{Bahr:2008pv}
M.~B{\"a}hr\hrefCMSnoop {}{ {et~al.}, ``Herwig++ physics and manual'',}
  \textit{ Eur. Phys. J. C} \textbf{ 58} (2008) 639,
  \href{http://dx.doi.org/10.1140/epjc/s10052-008-0798-9}{\doi{10.1140/epjc/s10052-008-0798-9}},
\href{http://www.arXiv.org/abs/0803.0883}{\texttt{arXiv:0803.0883}}.

\bibitem{Heister:2001jg}
\hrefCMSnoop {}{{ALEPH} Collaboration, ``{Study of the fragmentation of b
  quarks into B mesons at the Z peak}'',} \textit{ Phys. Lett. B} \textbf{ 512}
  (2001) 30,
  \href{http://dx.doi.org/10.1016/S0370-2693(01)00690-6}{\doi{10.1016/S0370-2693(01)00690-6}},
\href{http://www.arXiv.org/abs/hep-ex/0106051}{\texttt{arXiv:hep-ex/0106051}}.

\bibitem{DELPHI:2011aa}
\hrefCMSnoop {}{{DELPHI} Collaboration, ``{A study of the b-quark fragmentation
  function with the DELPHI detector at LEP I and an averaged distribution
  obtained at the Z pole}'',} \textit{ Eur. Phys. J. C} \textbf{ 71} (2011)
  1557,
  \href{http://dx.doi.org/10.1140/epjc/s10052-011-1557-x}{\doi{10.1140/epjc/s10052-011-1557-x}},
\href{http://www.arXiv.org/abs/1102.4748}{\texttt{arXiv:1102.4748}}.

\bibitem{Bowler:1981sb}
\hrefCMSnoop {}{M.~G. Bowler, ``${\rm e}^{+}{\rm e}^{-}$ {Production} of heavy
  quarks in the string model'',} \textit{ Z. Phys.} \textbf{ 11} (1981) 169,
\href{http://dx.doi.org/10.1007/BF01574001}{\doi{10.1007/BF01574001}}.

\bibitem{Peterson:1982ak}
\hrefCMSnoop {}{C.~Peterson, D.~Schlatter, I.~Schmitt, and P.~M. Zerwas,
  ``{Scaling violations in inclusive ${\rm e}^{+}{\rm e}^{-}$ annihilation
  spectra}'',} \textit{ Phys. Rev. D} \textbf{ 27} (1983) 105,
\href{http://dx.doi.org/10.1103/PhysRevD.27.105}{\doi{10.1103/PhysRevD.27.105}}.

\bibitem{Rojo:2015acz}
\hrefCMSnoop {}{J.~Rojo {et~al.}, ``{The PDF4LHC report on PDFs and LHC data:
  Results from Run I and preparation for Run II}'',} \textit{ J. Phys. G}
  \textbf{ 42} (2015) 103103,
  \href{http://dx.doi.org/10.1088/0954-3899/42/10/103103}{\doi{10.1088/0954-3899/42/10/103103}},
\href{http://www.arXiv.org/abs/1507.00556}{\texttt{arXiv:1507.00556}}.

\bibitem{Butterworth:2015oua}
\hrefCMSnoop {}{J.~Butterworth {et~al.}, ``{PDF4LHC recommendations for LHC Run
  II}'',} \textit{ J. Phys. G} \textbf{ 43} (2016) 023001,
  \href{http://dx.doi.org/10.1088/0954-3899/43/2/023001}{\doi{10.1088/0954-3899/43/2/023001}},
\href{http://www.arXiv.org/abs/1510.03865}{\texttt{arXiv:1510.03865}}.

\bibitem{Accardi:2016ndt}
\hrefCMSnoop {}{A.~Accardi {et~al.}, ``A critical appraisal and evaluation of
  modern {PDFs}'',} \textit{ Eur. Phys. J. C} \textbf{ 76} (2016) 471,
  \href{http://dx.doi.org/10.1140/epjc/s10052-016-4285-4}{\doi{10.1140/epjc/s10052-016-4285-4}},
\href{http://www.arXiv.org/abs/1603.08906}{\texttt{arXiv:1603.08906}}.

\bibitem{Czakon:2015owf}
\hrefCMSnoop {}{M.~Czakon, D.~Heymes, and A.~Mitov, ``{High-precision
  differential predictions for top-quark pairs at the LHC}'',} \textit{ Phys.
  Rev. Lett.} \textbf{ 116} (2016) 082003,
  \href{http://dx.doi.org/10.1103/PhysRevLett.116.082003}{\doi{10.1103/PhysRevLett.116.082003}},
\href{http://www.arXiv.org/abs/1511.00549}{\texttt{arXiv:1511.00549}}.

\bibitem{Khachatryan:2016mnb}
\hrefCMSnoop {}{{CMS Collaboration}, ``{Measurement of differential cross
  sections for top quark pair production using the lepton+jets final state in
  proton-proton collisions at 13 TeV}'',} \textit{ Phys. Rev. D} \textbf{ 95}
  (2017) 092001,
  \href{http://dx.doi.org/10.1103/PhysRevD.95.092001}{\doi{10.1103/PhysRevD.95.092001}},
\href{http://www.arXiv.org/abs/1610.04191}{\texttt{arXiv:1610.04191}}.

\bibitem{Sirunyan:2017mzl}
\hrefCMSnoop {}{{CMS Collaboration}, ``{Measurement of normalized differential
  \ttbar cross sections in the dilepton channel from pp collisions at $\sqrt{s}
  = 13$ TeV}'',} (2017).
  \href{http://www.arXiv.org/abs/1708.07638}{\texttt{arXiv:1708.07638}}.
Submitted to \textit{JHEP}.

\bibitem{Sirunyan:2018avv}
\hrefCMSnoop {}{{CMS Collaboration}, ``Study of the underlying event in top
  quark pair production in pp collisions at 13 {TeV}'',} (2018).
  \href{http://www.arXiv.org/abs/1807.02810}{\texttt{arXiv:1807.02810}}.
Submitted to \textit{Eur. Phys. J. C}.

\bibitem{Seymour:2013qka}
\hrefCMSnoop {}{M.~H. Seymour and A.~Siodmok, ``{Constraining MPI models using
  $\sigma_{eff}$ and recent Tevatron and LHC underlying event data}'',}
  \textit{ JHEP} \textbf{ 10} (2013) 113,
  \href{http://dx.doi.org/10.1007/JHEP10(2013)113}{\doi{10.1007/JHEP10(2013)113}},
\href{http://www.arXiv.org/abs/1307.5015}{\texttt{arXiv:1307.5015}}.

\bibitem{Sirunyan:2018wem}
\hrefCMSnoop {}{{CMS Collaboration}, ``Measurement of differential cross
  sections for the production of top quark pairs and of additional jets in
  lepton+jets events from pp collisions at $\sqrt{s} =$ 13 {TeV}'',} \textit{
  Phys. Rev. D} \textbf{ 97} (2018) 112003,
  \href{http://dx.doi.org/10.1103/PhysRevD.97.112003}{\doi{10.1103/PhysRevD.97.112003}},
\href{http://www.arXiv.org/abs/1803.08856}{\texttt{arXiv:1803.08856}}.

\bibitem{Alwall:2011uj}
J.~Alwall\hrefCMSnoop {}{ {et~al.}, ``{MadGraph 5: going beyond}'',} \textit{
  JHEP} \textbf{ 06} (2011) 128,
  \href{http://dx.doi.org/10.1007/JHEP06(2011)128}{\doi{10.1007/JHEP06(2011)128}},
\href{http://www.arXiv.org/abs/1106.0522}{\texttt{arXiv:1106.0522}}.

\bibitem{Field:2010bc}
\hrefCMSnoop {}{R.~Field, ``Early {LHC} underlying event data---findings and
  surprises'',} (2010).
\href{http://www.arXiv.org/abs/1010.3558}{\texttt{arXiv:1010.3558}}.

\end{thebibliography}\endgroup

\cleardoublepage \appendix\section{The CMS Collaboration \label{app:collab}}\begin{sloppypar}\hyphenpenalty=5000\widowpenalty=500\clubpenalty=5000\vskip\cmsinstskip
\textbf{Yerevan Physics Institute, Yerevan, Armenia}\\*[0pt]
A.M.~Sirunyan, A.~Tumasyan
\vskip\cmsinstskip
\textbf{Institut f\"{u}r Hochenergiephysik, Wien, Austria}\\*[0pt]
W.~Adam, F.~Ambrogi, E.~Asilar, T.~Bergauer, J.~Brandstetter, E.~Brondolin, M.~Dragicevic, J.~Er\"{o}, A.~Escalante~Del~Valle, M.~Flechl, M.~Friedl, R.~Fr\"{u}hwirth\cmsAuthorMark{1}, V.M.~Ghete, J.~Hrubec, M.~Jeitler\cmsAuthorMark{1}, N.~Krammer, I.~Kr\"{a}tschmer, D.~Liko, T.~Madlener, I.~Mikulec, N.~Rad, H.~Rohringer, J.~Schieck\cmsAuthorMark{1}, R.~Sch\"{o}fbeck, M.~Spanring, D.~Spitzbart, A.~Taurok, W.~Waltenberger, J.~Wittmann, C.-E.~Wulz\cmsAuthorMark{1}, M.~Zarucki
% ...
\vskip\cmsinstskip
% ...
1:  Also at Vienna University of Technology, Vienna, Austria\\
% ...
\end{sloppypar}
\end{document}